\documentclass[aps,prd,eqsecnum,twocolumn,twoside,showpacs,amssymb,nofootinbib]{revtex4}

\usepackage{amsfonts}
\usepackage{graphicx}
\usepackage{bm}
\begin{document}
\title{Localized Matter and  Geometry of the Dirac Field. }
\author{ Alexander Makhlin}
\email[]{amakhlin@comcast.net}
\affiliation{ Rapid Research, Inc., Southfield, MI, USA}
\date{October 11, 2009}
\pacs{11.27.+d; 11.30.Na}
\begin{abstract}

Within the framework of classical field theory, the connection
between the Dirac field as the field of matter and the spacetime
metric is discussed. Polarization structure of the Dirac field is
shown to be rich enough to determine the spacetime metric locally
and to explain the emergence of observed matter as localized
waveforms. The localization of the waveforms is explained as the
result of the local time slowdown and the Lorentz contraction as a
dynamic re-shaping of the waveforms in the course of their
acceleration. A definition of mass as a limiting curvature of the
spinor-induced metric is proposed. A view of the vacuum as a
uniformly distributed unit invariant density of the Dirac field
with an explicitly preserved invariance of the light cone is
brought forward. Qualitative explanation of the observed charge
asymmetry as the consequence of the dynamics of localization is
given. The classical pion field is obtained as a manifestation of
stresses, mass and charge flux in localized waveforms of the Dirac
field. Some implications of the finite size of colliding objects
for high-energy processes are discussed. A hypothesis that known
internal degrees of freedom are the local spacetime (angular)
coordinates that have no precise counterparts in Riemannian
geometry is proposed.

\end{abstract} \maketitle

\section{\normalsize Introduction}

The purpose of this paper is to show that the classical theory of
the Dirac field, considered as a primary form of matter can
explain those important properties of the observed matter which so
far remain a mystery when viewed from the perspective of quantum
field theory. In the first place, these properties are
localization of elementary objects and the origin of their mass
and finite size. Another, not less intriguing question is the
origin of the observed charge asymmetry of normal matter -- we
find only small, heavy, positively charged protons (nuclei) and
light, negatively charged, poorly localized electrons as the only
stable particles around us. It seems that a possibility to answer
these big questions has been overlooked at the early stage of
field theory.

An idea that the fields $\psi(x)$ of matter themselves can
immediately define the metric tensor $g_{\mu\nu}(x)$ was brought
forward by Wigner \cite{Wigner1} and Sakharov \cite{Sakharov1}.
From the physics perspective, this idea is extremely sound;
coordinates can be measured only through positions and shapes of
material bodies. [In quoted works, tensor indices of $g_{\mu\nu}$
were due to the derivatives $\partial_\mu\psi(x)$.]  It appears
that the Dirac field builds up the metric of spacetime without
resorting to {\em ad hoc} derivatives and it does this in such a
way that the time slows down in the domains of magnified invariant
density. This observation alone leads to a natural and startlingly
elegant answer to these big questions. Merely in the spirit of
Huygens principle, this fact leads to self-localization of the
Dirac wave forms into small, heavy, positively charged objects of
finite size while leaving a negatively charged fraction of matter
in the form of an agile substance surrounding these small and
heavy objects. It also changes the image of the Dirac sea as the
vacuum -- a uniformly distributed unit invariant density ${\cal
R}=1$ is identified with $g_{00}=1/{\cal R}^2=1$ and replaces a
continuum of oscillators with an unbound energy spectrum.

From the perspective of the present work, the Dirac field is
important, not as a special representation of the Lorentz group,
but as a field that accurately describes the hydrogen atom.
Lagrangian formalism is not used and no symmetry is assumed {\em a
priori}. The main focus is on the possibility of deriving the most
important properties of observed stable matter and its motion
starting from the basic properties of the Dirac field and its
equation of motion. No significant attempt to develop a formal
perturbation theory that could have dealt with the finite size of
particles as the {\it in}- and {\it out}- states of the quantum
scattering process has been made so far. Once localized wave forms
are found they immediately can be used as a basis for second
quantization and their fields can serve as Heisenberg operators.
The best prospect of this study is connected with the possibility
to bridge the gap between point-like particles of classical
electrodynamics and the plane waves of the quantum theory of
scattering.

The logic of the present work can be outlined as follows:

The stage is set in Sec.\ref{ssec:Sec2A}, beginning from a review
of well-known properties of the bilinear forms of the Dirac field
with emphasis on their purely algebraic origin. These forms are
empirically verified to be affine Lorentz tensors at a generic
point and they are further used to build a quadruple of orthogonal
Lorentz unit vectors (tetrad). The possibility of treating these
unit vectors as the tangent vectors of the coordinate lines of a
usual holonomic coordinate system and thus to define the
Riemannian metric as a descendant of the Dirac field is considered
in Sec.\ref{ssec:Sec2B}. It appears that certain conditions of
integrability should be met and that these conditions are
controlled by the Dirac equation. A key observation that the
matter-induced metric must be equivalent to a long range
interaction is made.

The rules of differential calculus for the Dirac field in curved
spacetime  are reviewed in Appendix \ref{app:appA}, mostly
following V. Fock \cite{Fock1}. A greater generality than in
\cite{Fock1} is intentionally admitted -- and there was no
possibility to truncate it later. Sec.\ref{sec:Sec3} thoroughly
investigates if various differential identities, derived from the
Dirac equation, can be put in the form of tensor equations, thus
being independent of a particular choice of coordinate system. For
identities that involve the energy-momentum $T^a_b$, the
conclusion is negative with the following facts firmly
established: (i) The normal covariant form of the energy-momentum
conservation cannot be assembled when the coordinate system is
normal. (ii) The tetrad components  $T^a_b$ of energy-momentum are
not the invariants of tensors. (iii) Being formally translated
into coordinate form, the identity of energy-momentum balance
keeps an explicit dependence on tetrad vectors. It does not
reproduce the equation for a geodesic line in a given metric
background. (iv) An explicit expression for the force of gravity
(inertia) is derived from the constraint, which accounts for the
interplay between scalar and pseudoscalar quantities in the course
of a physical acceleration of a Dirac object. Only in a crude
approximation of a point-like object is the standard metric
expression recovered. To account for the flux of momenta in
spacelike directions (pressure) in a localized waveform, a new
stress tensor, $P^a_b$, is introduced and studied in the same
detail in Appendix \ref{app:appB}. A connection with the theory of
Nambu and Jona-Lasinio is traced. The wave equation for the
pseudoscalar density (pion field) with the source that has the
structure of the axial anomaly is derived in Section
\ref{sapp:appB2} .

An unremovable dependence on tetrad vectors prompted a detailed
investigation (in Sec.\ref{sec:Sec4}) of the geometric properties
of vector and axial currents and constraints that affect the
integrability of differential equations for their lines. It is
shown that for the on-mass-shell Dirac field, the timelike
congruence of lines of the vector current always is normal so that
there always exists a system of hypersurfaces of a constant time.
The key result reads as $dt={\cal R} ds_0$, where ${\cal R}=
\sqrt{j^2}$ is the invariant density of the Dirac field; $dt$ and
$ds_0$ are the intervals of the world and proper time,
respectively. It immediately predicts a general trend of
self-localization for the Dirac field and a Lorentz contraction of
accelerated elementary objects as a physical process.
Investigation of constraints connected with identities for the
axial current (which, having a source, determines a radial
direction) brought about another result -- the maximal curvature
of the 2-d surface of constant radius cannot exceed $m$, the mass
parameter in the Dirac equation.  A set of relations that connect
bending of coordinate lines with the distribution of the axial
current is obtained. These relations prompt a strong parallel
between the local dynamics of the Dirac field and systems of
inertial navigation -- linear acceleration inevitably causes a
precession and {\em vice versa}. Sec.\ref{ssec:Sec4C} deals with
the intuitively appealing (and possibly not realistic) case of
normal radial coordinate, in which behavior of the angular
coordinates can be studied analytically.

With the metric explicitly depending on the field of matter the
Dirac equation becomes nonlinear in a unique way, which leads to
self-localization as an intrinsic property of the Dirac field.
Different forms of this equation along with an analysis of
individual terms are the subject of Sec.\ref{sec:Sec5} and
Appendix \ref{app:appC}.

In Sec.\ref{sec:Sec6}, a major conjecture regarding the nature of
electric charge is made. Maxwell equations are introduced and it
is shown that a stable Dirac waveform cannot interact with its own
electric field. Furthermore, two such forms cannot intersect each
other in spacetime. The origin of electromagnetic radiation is
explicitly traced back to the loss of simultaneity between the
Dirac waveform and its Coulomb field.

We conclude in Sec.\ref{sec:Sec7} with a short list of the
existing data and experiments that are in line with or can serve
as the tests for our predictions.

The results of this work, if looked at as a launch-pad for further
investigations, are striking in their anticipated mathematical
complexity and physical transparency. It seems, however, that the
former is the inevitable toll for the latter. The nonlinearity of
the Dirac equation makes finding its explicit solutions a
formidable task. But this nonlinearity is not artificial -- no
{\em ad hoc} nonlinear terms were added to the basic Dirac
Lagrangian in order to simulate any experimentally found patterns
of matter behavior, symmetry, etc. On the contrary, the discovered
generic structure corresponds to the perfectly understood
phenomenon of localization, which is due to the local time
slowdown, and then the loss of certain elements of spatial
symmetry due to localization. Despite being genuinely nonlinear,
these phenomena are so natural for any kind of wave propagation
that only a minimal amount of information about the physical
nature of the waves is needed to not only understand the whole
picture qualitatively, but even to make semi-quantitative
estimates.  In the text we also outline how the existence of the
pion field or how the known properties of the neutrino can be
inferred from the concept of a localized Dirac waveform.

\section{\normalsize Dirac field and Riemannian geometry.  \label{sec:Sec2}}

The first attempts to bring the Dirac equation into the framework
of General Relativity (GR) was made by V. Fock \cite{Fock1} and H.
Weyl \cite{Weyl} in a series of papers in 1929. This study (and
many other studies of that year) was in line with the basic
concept of Einstein's GR that, in the local limit (inertial
reference frame), one has to reproduce the results of  special
relativity; it was established earlier that spinors do indeed
provide a linear representation of the Lorentz group. Somewhat
later, E. Cartan pointed to a insurmountable difficulty -- there
are no representations of the general linear  group of
transformations $GL(4)$  that are similar to spinor
representations of the Lorentz group of rotations. Cartan stated
the following theorem , which {\em vetoed} spinors in Riemannian
geometry:

\textquotedblleft {\em  With the geometric sense given to the word
\textquotedblleft spinor" it is impossible to introduce spinors
into classical Riemannian technique; i.e., having chosen an
arbitrary system of co-ordinates $x^\mu$ for space, it is
impossible to represent spinor by any \underline{finite number} of
components $\psi_i$ such that $\psi_i$ have covariant derivatives
of the form $\psi_{i;\mu}=\partial_\mu\psi_i +
\Gamma^j_{i\mu}\psi_j$, where $\Gamma^j_{i\mu}$  are
\underline{determinate functions} of $x^\mu$. }" \cite{Cartan}

Of these two underscored reservations of Cartan, the first one was
investigated by Ne'eman {\em et al} \cite{Neeman}, who proposed to
overcome the veto by resorting to the infinite-dimensional
representations of the Lorentz group. The present study explores
the window, which is left open by the second reservation. As long
as natural coordinates for the Dirac field are anholonomic (also
in the sense of the theory of dynamical systems) the
\textquotedblleft connections" $\Gamma^j_{i\mu}$ and, eventually,
the metric tensor $g_{\mu\nu}(x)$  appear to be functions of the
Dirac field (which are determined by Eqs. (\ref{eq:E4.7}),
(\ref{eq:E4.11}), (\ref{eq:E4.15}) and (\ref{eq:E4.19}),
(\ref{eq:E4.20}) below) and not determinate functions of $x^\mu$.

In this section we set the stage by demonstrating that the two key
issues of geometry, direction and distance, can be separated in a
\textquotedblleft physical way" by associating the field of
directions with the Dirac field of matter. The Riemannian metric
will then be associated with the propagation of signals.

\subsection{\normalsize Algebraic properties of the Dirac Field.  \label{ssec:Sec2A}}

All observables associated with the Dirac field are  bilinear
forms built with the aid of Dirac matrices $\alpha^i$ and $\beta$,
which satisfy the commutation relations ( $\alpha^a=
(1,\alpha^i)$; $a=0,1,2,3$; $i=1,2,3$)
\begin{equation}\label{eq:E2.1}
\alpha^a \beta \alpha^b  +  \alpha^b \beta \alpha^a= 2\beta
\eta^{ab}~,
\end{equation}
where $ \eta^{ab} = {\rm diag}(1,-1,-1,-1)$. We begin with a
review of the properties of the Dirac field $\psi(x)$ which hold
at {\em a point}, without a precise definition of the coordinates
$x$. For now, $\psi$ will stand for a column of four complex
numbers $\psi_\sigma$.

There are sixteen linearly independent $4\times 4$ Hermitian
matrices all of which can be constructed from the four matrices
$\alpha^i$ and $\beta$. The Dirac matrices,  $~\rho_i~$
($~\rho_1=\beta$, $\rho_3=-i\alpha^1\alpha^2\alpha^3$, $~\rho_2
=-i\beta\rho_3$), and $~\sigma_i=\rho_{3}\alpha^i ~$ satisfy the
same commutation relations as the Pauli matrices, and all $\sigma$
matrices commute with the $\rho$ matrices:~
$\sigma_i\sigma_k=\delta_{ik}+i\epsilon_{ikl}\sigma_k$,
$\rho_a\rho_b =\delta_{ab}+i\epsilon_{abc}\rho_c$, $\sigma_i\rho_a
-\rho_a\sigma_i =0$. The matrices $-\rho_3$ and  $\rho_1$ are
commonly known as $\gamma^5$ and $\gamma^0$, respectively
\footnote{We consciously refrain from using the anti-hermitian
matrices $\gamma^i=\rho_1\alpha^i$ and the Pauli-conjugated
spinors $\bar{\psi}=\psi^+\rho_1$. In their terms, the formulae of
parallel transport (see Appendix A) would be much less transparent
and unnecessarily complicated.}. Below, these matrices are used in
the spinor representation,
\begin{eqnarray}
\alpha_i=\left( \begin{array}{c c} \tau_i  &  0 \\ 0&
-\tau_i\end{array}\right),~~~~~~ \rho_1=\left( \begin{array}{c c}
0  & {\bf 1}  \\
 {\bf 1}& 0 \end{array}\right),\nonumber\\
\rho_2=\left( \begin{array}{c c} 0 &  -i\cdot{\bf 1}  \\
 i\cdot{\bf 1} & 0 \end{array} \right),~~
 \rho_3=\left( \begin{array}{c c} {\bf 1}  &  0  \\
  0& -{\bf 1} \end{array} \right).\nonumber
\end{eqnarray}
where   $\tau_i$ are the $2\times 2$ Pauli matrices. Employing the
Dirac matrices, we can define the four components of the
\textquotedblleft vector current",
$j^a=\psi^+\alpha^a\psi\equiv\bar{\psi}\gamma^a\psi$, the four
components of the \textquotedblleft axial current", ${\cal J}^a=
\psi^+\rho_3\alpha^a\psi\equiv\bar{\psi}\gamma^5\gamma^a\psi $,
the \textquotedblleft scalar" ${\cal S}
=\psi^+\rho_1\psi\equiv\bar{\psi}\psi$ and \textquotedblleft
pseudoscalar"${\cal P} =\psi^+\rho_2\psi\equiv
-i\bar{\psi}\gamma^5\psi$, and the six components of the
skew-symmetric \textquotedblleft tensor" ${\cal M}^{ab}=
(i/2)\psi^+[\alpha^a \rho_1 \alpha^b- \alpha^b\rho_1\alpha^a]
\psi$. The similarity of these quantities to the Lorentz tensors
can be verified in a purely algebraic way. Indeed, if the Dirac
field $\psi$ is transformed by means of a substitution $\psi \to
S\psi$ (or the matrices are transformed as $\alpha^a\to
S^+\alpha^aS$, etc.), with the matrix $S$ depending on four
complex parameters,
\begin{eqnarray} \label{eq:E2.2}
S=\left(\begin{array}{c c}\lambda & 0\\0 &
(\lambda^+)^{-1}\end{array}\right); ~~~~~~ \lambda=\left(
\begin{array}{c c}\alpha  & \beta
\\ \gamma & \delta \end{array} \right),  \nonumber \\
{\rm det}|\lambda|\equiv \alpha\delta-\beta\gamma=1~,~~~~~
\end{eqnarray}
then the components of $j^a$, ${\cal J}^a$, ${\cal S}$, ${\cal P}$
and ${\cal M}^{ab}$ experience a four-dimensional Lorentz rotation
at angles which are uniquely determined by these parameters. For
example, if we take $\alpha=e^{-i\phi/2}$, $\beta=\gamma=0$, and
$\delta=e^{i\phi/2}$, then the transformation
$S=e^{-i\phi\sigma_3/2}$ is unitary and the components of
$j'^a=\psi^+S^+\alpha^aS\psi$ are
\begin{eqnarray}\label{eq:E2.3}
j'^0=j^0,~~j'^1=j^1\cos\phi-j^2\sin\phi,~~  \nonumber \\
j'^2=j^1\sin\phi+j^2\cos\phi,~~j'^3=j^3,
\end{eqnarray}
which corresponds to the rotation of the vector $j^a$ at an angle
$\phi$ around the axis  \textquotedblleft 3". In exactly the same
way, if we take $\alpha=e^{-\eta/2}$, $\beta=\gamma=0$, and
$\delta=e^{\eta/2}$ then $S=e^{-\eta\alpha_3/2}$; the components
of   $j'^a$ will be
\begin{eqnarray}\label{eq:E2.4}
j'^0=j^0\cosh\eta-j^3\sinh\eta,~~j'^1=j^1,~~   \nonumber \\
j'^2=j^2,~~j'^3=-j^0\sinh\eta+j^3\cosh\eta.
\end{eqnarray}
This  transformation of $\psi$ corresponds to a Lorentz boost in
the  \textquotedblleft third" direction and is {\em not unitary}.
Similar correct relations are immediately verified for the scalars
${\cal S}'$ and ${\cal P}'$, the  vector ${\cal J}'^a=
\psi^+S^+\rho_3 \alpha^aS\psi$, etc. Therefore, for example, the
quantities
\begin{eqnarray}\label{eq:E2.5}
j^2=\eta_{ab}j^aj^b=j_0^2-{\vec j}^2,~~~ {\cal J}^2=\eta_{ab}{\cal
J}^a {\cal J}^b, \nonumber \\ j\cdot{\cal J}=\eta_{ab}j^a{\cal
J}^b=j_0{\cal J}_0- {\vec j}{\vec{\cal J}}~~~~~~~~~~
\end{eqnarray}
are invariants of the $\psi \to S\psi$ transformations. Notably,
the Minkowski signature matrix of Eq.(\ref{eq:E2.1}),
$\eta^{ab}\equiv \eta_{(a)}\delta^a_b$, and its inverse
$\eta_{ab}$, $\eta^{ab} \eta_{bc}=\delta^a_c$, came up here in a
purely algebraic way, without even mentioning the Lorentz symmetry
and we will use it right away in order to preserve the usual
convention about contraction of repeated  upper and lower indices.
It is also just an algebraic exercise to check that ${\cal
R}^2\equiv j^aj_a =-{\cal J}^a{\cal J}_a={\cal S}^2+{\cal P}^2>0$
and that $j \cdot {\cal J}=0$.  The latter relation means that if
the vector current of the transformed field is of the form
$j^a=({\cal R},{\vec 0})$ then the axial current can only be of
the form ${\cal J}^a=(0,{\vec{\cal J}})$ and that it can be
further \textquotedblleft rotated" to ${\cal J}^a=(0,0,0,{\cal
R})$. Therefore, at a generic point $x$, the vectors
$e^a_{(0)}(x)=j^a/{\cal R}$ and $e^a_{(3)}(x)= {\cal J}^a/{\cal
R}$ are the orthogonal timelike and spacelike unit vectors,
respectively, and they can be reduced to
$e^a_{(0)}\circeq\delta^a_0$ and $e^a_{(3)}\circeq\delta^a_3$. (In
what follows, the symbol $(\circeq)$ is used in equations that
imply such a particular reduction.)

The components of the tensor ${\cal M}^{ab}$ and its dual
${\overstar{\cal M}}^{ab}=(1/2)\epsilon^{abcd}{\cal M}_{cd}$ are
\begin{eqnarray}\label{eq:E2.6}
{\cal M}^{0i} = K_i=\psi^+\rho_2\sigma_i\psi,~~~ {\overstar{\cal
M}}^{ij}= \epsilon^{0ijm}K_m    \nonumber  \\ {\overstar{\cal
M}}^{0i}= L_i=\psi^+\rho_1\sigma_i\psi,~~~ {\cal M}^{ij} =
\epsilon^{0ijm}L_m.
\end{eqnarray}
Because ${\cal M}^{ab}{\overstar{\cal M}}_{bc}
=(\vec{L}\cdot\vec{K})\delta^a_c$, these two tensors can be used
to build two couples of vectors which are spacelike, orthogonal to
$e^a_{(0)}$ and $e^a_{(3)}$ and to each other,
\begin{eqnarray} \label{eq:E2.7}
E_c=(j^a/ {\cal R}) {\cal M}_{ab}[\delta^b_c +{\cal J}^b{\cal J}_c
/{\cal R}^2]\circeq(0,K_1,K_2,0) ,~~~  \nonumber       \\
{{\overstar E}_c}=({\cal J}^a/{\cal R}) {\overstar{\cal M}}_{ab}
[\delta^b_c- j^b j_c/{\cal R}^2]\circeq(0,K_2,-K_1,0),~\nonumber
\\ H_c=({\cal J}^a/ {\cal R}) {\cal M }_{ab} [\delta^b_c-j^b
j_c/{\cal R}^2]\circeq(0,-L_2,L_1,0) , ~~\nonumber      \\
{\overstar H}_c=(j^a/ {\cal R}) {\overstar{\cal M}}_{ab}
[\delta^b_c+ {\cal J}^b {\cal J}_c/{\cal R}^2]
\circeq(0,L_1,L_2,0).~~~~
\end{eqnarray}

A full set of easily verifiable {\em identities} between
invariants of the transformations (\ref{eq:E2.2}) is given by
\begin{eqnarray}\label{eq:E2.8}
{\cal R}^2\equiv j_a j^a=-{\cal J}_a {\cal J}^{a}={\cal S}^2+
{\cal P}^2, ~~{\cal J}_a j^{a}=0,\nonumber\\     {\cal S}^2 -
{\cal P}^2=\vec{L}^2 -\vec{K}^2,~~~ {\cal S}{\cal
P}=\vec{L}\cdot\vec{K}.~~~~~~~~
\end{eqnarray}
The scalars allow for the following parameterizations,
\begin{eqnarray}\label{eq:E2.9}
{\cal S}={\cal R}\cos\Upsilon~,~~~  {\cal P}={\cal
R}\sin\Upsilon~,
\end{eqnarray}
where both ${\cal R}$ and $\Upsilon $ are functions of the Dirac
field. It is important that the absolute values of ${\cal S}$ and
${\cal P}$ do not exceed ${\cal R}$. A similar observation is true
for the second line of Eq.(\ref{eq:E2.8}),
\begin{eqnarray}\label{eq:E2.10}
{\cal S}^2-{\cal P}^2=\vec{L}^2 -\vec{K}^2= {\cal R}^2 \cos
2\Upsilon, \nonumber\\  2{\cal S}{\cal P}
=2\vec{L}\cdot\vec{K}={\cal R}^2\sin 2\Upsilon.
\end{eqnarray}

Concluding the discussion of the algebraic properties of bilinear
forms  of the Dirac field at a generic point $x$, let us
introduce, along with the orthogonal system $e^\beta_{(a)}[\psi]$,
a reciprocal (in algebraic sense) system $e_\beta^{(a)}[\psi]$
\begin{eqnarray}\label{eq:E2.12}
\sum_\alpha e^\alpha_{(a)}e_\alpha^{(b)}=\delta_a^b, ~~~ \sum_a
e^\alpha_{(a)}e_\beta^{(a)}=\delta_\beta^\alpha,~~
\end{eqnarray}
where we  assumed that ${\rm det}|e^\beta_{(a)}|\neq 0$. Then, a
simple algebra verifies that the objects
\begin{eqnarray}\label{eq:E2.13}
g_{\alpha\beta}= \eta_{ab}e_\alpha^{(a)}  e_\beta^{(b)},~~~
g^{\alpha\beta}= \eta^{ab}e^\alpha_{(a)}  e^\beta_{(b)}~~~
\end{eqnarray}
can be used to move the Greek indices up and down, for example, $$
g_{\alpha\beta} e^\beta_{(b)}=  e_{(a)\alpha}e_\beta^{(a)}
e^\beta_{(b)} =  \delta^a_b e_{(a)\alpha}=e_{\alpha(b)}. $$ It is
also evident that the repeated upper and lower Greek indices are
contracted.

\subsection{\normalsize Dirac currents and Riemannian geometry.  \label{ssec:Sec2B} }

From now on, we look at the $\psi_\sigma(x)$ as the physical Dirac
field, the continuous functions of the arbitrarily parameterized
points $x^\mu=(x^0,x^1,x^2,x^3)$ of the spacetime. So far, we have
verified that the algebraic structure of bilinear forms of the
Dirac field naturally contains an orthogonal quadruple of unit
Lorentz vectors at a generic point, thus defining spacetime
directions {\em at that point}. In a sense, linear transformations
(\ref{eq:E2.2}) of the Dirac field generate a group of homogeneous
linear transformations for vectors, thus associating with every
point a local centered affine space. Vectors of this quadruple are
thought of as smooth functions of $\psi(x)$ and it is tempting to
immediately treat it as a quadruple of the vector fields.  But
these transformations of the Dirac field have nothing to do with
the general transformation of coordinates, which are arguments of
$\psi(x)$. For a given fixed $\lambda$, we can consider
$x^\lambda= const$ as the equation of a coordinate hypersurface
and the lines along which all coordinates but $x^\lambda$ are
constant as coordinate lines. Tangent vectors of these lines
(which are gradients of the linear function $\varphi(x)=
x^\lambda$) are $h_{(\lambda)}^\mu=\partial x^\mu/\partial
x^\lambda= \delta_{(\lambda)}^\mu$; therefore, this coordinate
system is an holonomic one, but it has no metric and there is no
way to determine if its coordinate lines are orthogonal. One may
replace $x^\mu$ by smooth functions of other coordinates $y^\mu$,
$x^\mu=f^\mu(y)$, thus redefining coordinate lines and surfaces,
but such a change does not alter $\psi(x(y))$ and has nothing to
do with affine Lorentz transformations (\ref{eq:E2.2}). To bridge
the gap between the abstract field of directions determined by the
Dirac field and the given above definition of the holonomic
coordinates, it is necessary to know in advance that four systems
of differential equations (for the unknown $x^\mu$),
\begin{equation}\label{eq:E2.14}
{dx^0\over e^0_{(a)}(x)}= {dx^1\over e^1_{(a)}(x)}={dx^2\over
e^2_{(a)}(x)}={dx^3\over e^3_{(a)}(x)}= ds^a ~,
\end{equation}
for congruences of lines (labeled by the ordinal numbers $(a)$ )
are solvable and thus determine a coordinate net. In other words,
if these equations are integrable, then the system
\begin{equation}\label{eq:E2.15}
dx^\alpha = e^\alpha_{(a)}ds_a~, ~~~\mu=0,1,2,3,
\end{equation}
will represent lines, which at every point $x$ have a determinate
direction $e^\alpha_{(a)}(\psi)$, and only one line of the
congruence $(a)$ passes though each point in spacetime. The tetrad
$e^\alpha_{(a)}$ will be a Lorentz vector and a coordinate vector.
A change of the vector  variables $x^\mu=f^\mu(y)$ in
Eq.(\ref{eq:E2.15}) will result in the transformation of the
differential $dx^\mu$, $$dx^\alpha={\partial x^\alpha\over\partial
y^\sigma}dy^\sigma, ~~ {\partial x^\alpha\over\partial
y^\sigma}\cdot{\partial y^\sigma \over\partial x^\beta
}=\delta^\alpha_\beta,$$ and the equation for the same congruence
in new coordinates will read as
\begin{equation}\label{eq:E2.16}
dy^\sigma=e^\sigma_{(a)}(y)ds_a ,~~~~
e^\sigma_{(a)}(y)\stackrel{\rm def}{=} {\partial
y^\sigma\over\partial x^\beta} \cdot e^\beta_{(a)}[\psi].
\end{equation}

A new element here is that tangent vectors depend on the
coordinates via the coordinate dependence of the Dirac field
which, in its turn, is constrained by the equations of motion.
Therefore, the problem of integrability of Eqs.(\ref{eq:E2.14}) -
(\ref{eq:E2.16}) cannot be addressed solely within Riemannian
geometry; at least some properties of congruences must be
controlled by the Dirac equation. For \textquotedblleft temporal"
and  \textquotedblleft radial" congruences, the Dirac equation
indeed yields a set of constraints with a clear physical meaning.
The properties of congruences of angular arcs (including their
symmetry), in general, not only explicitly depend on particular
solutions of the Dirac equation but there may even be no
meaningful holonomic coordinates associated with these arcs.
Nevertheless, even keeping such a difficult perspective in mind,
let us consider all four tetrad vectors as contravariant vectors
of Riemannian geometry.

Traditional approaches assume solving the Dirac equation in a
determinate metric field $g_{\mu\nu}(x)$ of spacetime. The
polarization properties of the Dirac field prompt the opposite
direction of thinking. Namely, the field $\psi(x)$ must be the
solution of the Dirac equation, which explicitly depends on a
resulting metric $g_{\mu\nu}[\psi(x)]$ given by
Eqs.(\ref{eq:E2.13}). In such a context, the Minkowski form of the
metric in the local limit is associated not with an imaginable
local inertial frame but rather with the algebraic properties of
the Dirac field and (complementary to the latter) the hyperbolic
character of the Dirac equation. Thus, it is possible to overcome
Cartan's veto in two major points. First, there is no arbitrary
coordinates for spacetime (modulo a trivial change of variables).
Second, the connections, $\Gamma$, are no longer determinate
functions of $x$; they become functions of the Dirac field. In
this framework, as is shown below, the hypersurfaces of a constant
temporal coordinate naturally emerge; their existence is a
prerequisite for the quantization of the on-mass-shell Dirac field
in curved spacetime.  The proper time slows down in domains of a
higher matter density, which points to self-localization as an
intrinsic property of the Dirac field. This effect also clarifies
the nature of electric charge and of charge asymmetry of the
empirically known stable matter. Along with localized matter,
there always exists a preferred system of orthogonal congruences
determined by the internal polarization structure of physical
objects. In general, it can be considered as {\em the net of the
anholonomic coordinate system} \cite{Schouten}.

Let us follow the key idea of intrinsic geometry to associate
tensor fields with mutual invariants of tensors and parameters
(tangent vectors $e^\mu_{(a)}(x)$) of a system of congruences.
Furthermore, let us read Eqs.(\ref{eq:E2.12}) and (\ref{eq:E2.13})
as
\begin{eqnarray}\label{eq:E2.17}
g_{\nu\mu}(x) e^\mu_{(a)}(x)e^\nu_{(b)}(x)=\eta_{ab} \nonumber \\
\eta_{ab} e_\mu^{(a)}(x)e_\nu^{(b)}(x)=g_{\nu\mu}(x),
\end{eqnarray}
and consider {\em this} $g_{\nu\mu}(x)$, as a primary choice of
the spacetime metric. Only by virtue of  Eqs.(\ref{eq:E2.14}) can
we translate the first of equations Eq.(\ref{eq:E2.17}) into
\begin{eqnarray}\label{eq:E2.18}
ds^2=\eta_{ab}ds^a ds^b=g_{\nu\mu}(x)
e^\mu_{(a)}(x)e^\nu_{(b)}(x)ds^a ds^b \nonumber \\ =
g_{\nu\mu}(x)dx^\mu dx^\nu,~~~~~~~~~~~~~~~
\end{eqnarray}
and reconcile  Eqs.(\ref{eq:E2.13}) (inspired by the algebra of
the Dirac matrices) with the measure of length postulated in
Riemannian geometry.

When  $e_\mu^{(a)}$ is a vector with the law of transformation
(\ref{eq:E2.16}) and $g_{\nu\mu}(x)$ is a tensor (not necessarily
determining a metric) then the covariant derivative $\nabla_\nu
e_\mu^{(a)}$ with respect to $g_{\nu\mu}$ is also a tensor
\cite{Eisenhart}. Therefore, one can introduce a system of
invariants (the Ricci coefficients of rotation of a system of
congruences)
\begin{equation}\label{eq:E2.19}
\omega_{bca}=e_{(a)}^\mu(\nabla_\mu e_{(b)}^{\nu})
e_{(c)\nu}=-\omega_{cba}~.
\end{equation}
For a given $(c)$, six parameters $\omega_{abc}ds$ determine an
infinitesimal rotation of the pyramid of tetrad vectors in the
\textquotedblleft plane" $(ab)$ when the vertex of the pyramid is
displaced by $ds$ along a line of congruence $(c)$. Equation
\begin{equation}\label{eq:E2.20}
\nabla_\mu e_{(b)\nu}=\omega_{bca} e^{(c)}_{\nu}e^{(a)}_\mu
\end{equation}
is the inverse of (\ref{eq:E2.19}). Using Eqs.(\ref{eq:E2.19}) and
(\ref{eq:E2.20}), it is straightforward to check that if
$g_{\nu\mu}(x)$ has the form (\ref{eq:E2.17}) then $\nabla_\lambda
g_{\nu\mu}=0$. Consequently, the vector connections
$\Gamma^\sigma_{\mu\nu}$ coincide with the Christoffel symbols of
the metric $g_{\nu\mu}$.

In an ideal geometric world (i.e. when all four holonomic
coordinates exist) the necessary conditions for integrability of
Eqs.(\ref{eq:E2.20}) are given by the Ricci identities
\cite{Eisenhart},
\begin{eqnarray}\label{eq:E2.21}
(\nabla_\mu  \nabla_\lambda - \nabla_\lambda\nabla_\mu)
e_{(a)\nu} =e_{(a)}^\sigma R_{\sigma\nu\mu\lambda}
\end{eqnarray}
where $R_{\sigma\nu\mu\lambda}$ is the Riemann curvature tensor.
These equations can be cast in the form,
\begin{eqnarray}\label{eq:E2.22}
e_{(a)}^\sigma e_{(b)}^\nu e_{(c)}^\mu e_{(d)}^\lambda
R_{\sigma\nu\mu\lambda} = R_{abcd},
\end{eqnarray}
where
\begin{eqnarray}\label{eq:E2.23}
R_{abcd}\equiv \partial_d\omega_{abc}-
\partial_c\omega_{abd}~~~~~~~~~~~~~~~~~~~ \\ +
\sum_f\eta_f[\omega_{fad}\omega_{fbc}- \omega_{fac}\omega_{fbd}+
\omega_{abf}(\omega_{fcd}-\omega_{fdc})], \nonumber
\end{eqnarray}
is a system of invariants, which is then known as the tetrad
representation of the Riemann tensor. Since at least some of the
Ricci coefficients of rotation will appear to be functions of the
Dirac field, this dependence will be carried through onto the
Riemann and Ricci tensors. The Einstein equations for the metric
field $g_{\nu\mu}(x)$ that describes motion of macroscopic objects
may appear to be descendants of the constraints stemming from the
Dirac equation.

To summarize, if in spacetime, with arbitrarily chosen holonomic
coordinates, $x^\mu$, the Dirac field $\psi(x)$ is defined and at
each point the 16 quantities, $e^\nu_{(a)}[\psi(x)]$, are computed
(along with the algebraically reciprocal system
$e_\nu^{(a)}[\psi(x)]$) then the metric $g_{\nu\mu}(x)$ of
spacetime is given by Eq.(\ref{eq:E2.17}) and the interval by
Eq.(\ref{eq:E2.18}). This metric depends on the Dirac field and is
not defined {\em a priori}. From the physical perspective, its
existence seems to be a privilege of  exceptional solutions rather
than a rule.

It is important to realize that the material Dirac field defines a
system of the {\em unit} vector fields $e_{(a)}^\mu(x)$ --
therefore, {\em the effect of such a matter-induced metric should
be equivalent to a long-range interaction between localized
objects}.

\section{\normalsize Differential identities for tensors.  \label{sec:Sec3}}

In order to find limitations on the metric of spacetime, which can
host the localized configurations of the Dirac field, we begin
with the examination of various identities that are consequences
of the Dirac equation. The question is, whether differentials of
various bilinear forms of the Dirac field, which are considered as
the physical observables, can be translated into covariant
derivatives of tensors. We use this question as a test of the
roots of the discrepancies that could have led to Cartan's veto.
It appears that these discrepancies correspond to the clearly
understood physical processes.

Following Fock and Weyl, we postulate that the equation of motion
of the Dirac field and its conjugate are
\begin{eqnarray}\label{eq:E3.1}
\alpha^a D_a\psi=-im\rho_1\psi,\\ \label{eq:E3.2}
\psi^+{\overleftarrow D}^+_a \alpha^a =im\psi^+\rho_1,
\end{eqnarray}
where the covariant derivative $D_a\psi=(\partial_a-\Gamma_a)\psi$
of the Dirac field is defined in Appendix \ref{app:appA}. The
object $D_\mu\psi=e^a_\mu D_a \psi= (\partial_\mu-\Gamma_\mu)\psi$
will be used only as a symbol, since it has no clear geometrical
meaning. The mass parameter in these equation is {\em a priori}
arbitrary. Because the Dirac field has the property of
self-localization, every stable localized waveform will determine
the corresponding value of $m$.

\subsection{Identities for  vector and axial currents.}

From the equations of motion (\ref{eq:E3.1}) and (\ref{eq:E3.2})
one immediately derives two well-known identities.   One of them,
\begin{eqnarray}\label{eq:E3.3}
D_a j^a=\nabla_\mu j^\mu={1\over\sqrt{-g}}
\partial_\mu[\sqrt{-g}\psi^+ \alpha^\mu \psi]=0,
\end{eqnarray}
clearly indicates conservation of the vector (probability) current
of the on-mass-shell Dirac field, while the second one indicates
that the axial current is not conserved,
\begin{eqnarray}\label{eq:E3.4}
D_a{\cal J}^a= \nabla_\mu {\cal J}^\mu=2m{\cal P},
\end{eqnarray}
and has the pseudoscalar density as a source.  The significance of
Eq.(\ref{eq:E3.4}) is due to the pseudoscalar density ${\cal P}$
on the r.h.s. Since ${\cal P}$ is localized not less than ${\cal
R}$ and the vector ${\cal J}^\mu$ is spacelike, the unit axial
vector $e_{(3)}^\mu$  defines the outward radial direction. The
existence of such a direction is a distinct characteristic of a
localized object.


\subsection{ Flux of momenta: not tensors.}

Consider now a more complicated object $T^a_b=i\psi^+ \alpha^a D_b
\psi$, the Hermitian part of which is  normally regarded as the
energy-momentum tensor of the Dirac field\footnote{The reader
should not be confused by how the standard name \textquotedblleft
energy-momentum tensor" is used. In the context of the present
work, the invariants $T^a_b$ and $P^a_b$ are the auxiliary
objects. We are interested only in identities that can be derived
from the Dirac equation in tetrad form and then translated, if
possible, into the tensor form. Only Hermitian part of these
objects enters the equations that allow for a physical
interpretation.}. Its components are interpreted as the flux of
components $iD_b$ of the momentum in the direction of congruence
of lines of the vector current $j^a$. Because the vector current
is {\em timelike}, this tensor is well-suited to describe the flux
of momenta carried by massive particles. The Lagrangian $L_D$ of
the Dirac field vanishes for the on-mass-shell configurations and
$T^a_b$ does not have a diagonal term, $-L_D\delta^a_b$, which
could have been responsible for the flux of momentum in the
spacelike direction (e.g., the pressure). Since the spacelike
radial direction is controlled by the axial current, we are led to
consider another object, the stress tensor $P^a_b=i\psi^+\rho_3
\alpha^a D_b \psi$, which accounts for the flux of momenta in the
radial direction. For stable localized wave forms, there must be
no flux of any observables in the spacelike outward direction.
However, if we decide to investigate a particle's Lorentz
contraction as a dynamic process or the decay of a long-lived
waveform (considered as a particle), then we are led to consider
the spacelike flux of momenta due to the \textquotedblleft phase
shifts" inside the wave form. Regardless of how adequate this
intuitive physical interpretation of $T^a_b$ or $P^a_b$ is, or
even without any physical interpretation, they both can be used to
derive various useful identities , which allow one to compute the
rotation coefficients $\omega_{abc}$ as functions of the Dirac
field and thus constrain the possible metric (\ref{eq:E2.18}). In
this section, we study $T^a_b$ in detail. The stress tensor
$P^a_b$ is studied in Appendix B along the same guidelines.
\begin{widetext}
One would expect that the absolute differential of $T_{ab}$, being
computed according to the Leibnitz rule, will be as follows,
\begin{eqnarray}\label{eq:E3.5}
D_c T_{ab}=\partial_c T_{ab}-\omega_{adc}T_{db}-
\omega_{bdc}T_{ad}\equiv\nabla_c T_{ab}~.
\end{eqnarray}
If this expectation turns out to be  justified then the usual
covariant derivative will be immediately reproduced as
\begin{eqnarray}\label{eq:E3.6}
\partial_\lambda
T_{\sigma\mu}\!-\Gamma^\nu_{\sigma\lambda}T_{\nu\mu}
\!-\Gamma^\nu_{\mu\lambda}T_{\sigma\nu}\! =\!e_\lambda^c
e_\sigma^a e_\mu^b \nabla_c T_{ab}\!=\!\nabla_\lambda
T_{\sigma\mu}.~~
\end{eqnarray}
Contrary to the expectation of (\ref{eq:E3.5}), the answer reads
\begin{eqnarray}\label{eq:E3.7}
D_c[\psi^+ \alpha^a {\overrightarrow D}_b \psi]=
\partial_c[\psi^+ \alpha^a {\overrightarrow D}_b \psi]
-\psi^+[\Gamma^+_c \alpha^a+\alpha^a \Gamma_c] {\overrightarrow
D}_b \psi= \partial_c[\psi^+ \alpha^a {\overrightarrow D}_b \psi]-
\omega_{adc}\psi^+ \alpha^d {\overrightarrow D}_b \psi,
\end{eqnarray}
with the last term of Eq.(\ref{eq:E3.5}) missing, and  no hope to
recover the full expression (\ref{eq:E3.6}) of the covariant
derivative of the tensor! Contracting indices $a$ and $c$ we
arrive at the expression,

\begin{eqnarray}\label{eq:E3.8}
D_a[\psi^+ \alpha^a {\overrightarrow D}_b \psi]= \partial_a[\psi^+
\alpha^a {\overrightarrow D}_b \psi]+ \omega_{acc}\psi^+ \alpha^a
{\overrightarrow D}_b  \psi = {1\over\sqrt{-g}}
{\partial\over\partial x^\nu}\bigg[\sqrt{-g} e^\nu_{(a)} ~(\psi^+
\alpha^a {\overrightarrow D}_b \psi) \bigg],
\end{eqnarray}
which is exactly what one may wish to have as the l.h.s. of a
conservation law for the energy-momentum of the Dirac field . The
missing term is exactly the one that does not let one interpret
equations like $\nabla_\sigma T^\sigma_\mu=0$ as conservation of
anything. However, at the moment, a covariance can not yet be
explicitly visible; it may occur that  the r.h.s. of an expected
conservation law recovers the covariance of the resulting identity
as a whole. The r.h.s. must be determined using the equations of
motion. Let us first rewrite the l.h.s. of (\ref{eq:E3.8}) as
\begin{eqnarray}
D_a[\psi^+ \alpha^a {\overrightarrow D}_b \psi]=\psi^+ \alpha^a
[{\overrightarrow D}_a {\overrightarrow D}_b   - {\overrightarrow
D}_b {\overrightarrow D}_a] \psi + \psi^+{\overleftarrow
D}^+_a\alpha^a{\overrightarrow D}_b \psi  + D_b(\psi^+ \alpha^a
{\overrightarrow D}_a \psi) - \psi^+{\overleftarrow
D}^+_b\alpha^a{\overrightarrow D}_a \psi.   \nonumber
\end{eqnarray}
\end{widetext}
By virtue of the equations of motion (and due to the Leibnitz
rule) the last three  terms exactly cancel out and the final
result is
\begin{eqnarray}\label{eq:E3.9}
D_a[i\psi^+ \alpha^a {\overrightarrow D}_b \psi] =i\psi^+ \alpha^a
[{\overrightarrow D}_a{\overrightarrow D}_b -{\overrightarrow D}_b
{\overrightarrow D}_a] \psi.
\end{eqnarray}
Using Eqs.(\ref{eq:A.10}) and (\ref{eq:A.11}) to separate the
terms with and without derivatives  in the commutator and
comparing with (\ref{eq:E3.8}) we find that the Dirac equation
yields the identity:
\begin{eqnarray}\label{eq:E3.12}
\partial_a  T_{ab}\!-\omega_{cac}T_{ab}
= \omega_{bca} T_{ac}\!-\omega_{cab} T_{ac}\! -i\psi^+\alpha^a
\mathbb{D}_{ab}\psi,\nonumber\\
\end{eqnarray}
where the commutator $\mathbb{D}_{ab}$ does not contain
derivatives of $\psi$. After moving the term $ \omega_{bca}
T_{ac}$ from the right to the left, the l.h.s. becomes, according
to (\ref{eq:E3.5}) and (\ref{eq:E3.6}), the system of mutual
invariants of a usual covariant divergence of the tensor
$T^\sigma_\mu$ and congruences $e^\mu_a$. It can be transformed
either into coordinate form (\ref{eq:E3.6}), which has no explicit
dependence on tetrad vectors or into a non-coordinate form.
Unfortunately, this coordinate independence of a fragment of
identity (\ref{eq:E3.12}) is useless, because there remains an
abnormal term $ \omega_{cab} T_{ac}$ on the right. Being
translated into a coordinate form, this term  becomes
$(\nabla_\lambda e^\nu_a)e^a_\sigma T^\sigma_\nu$. It explicitly
depends on tetrad vectors (on how the coordinate lines are
bending).

The abnormal term $ \omega_{cab} T_{ac}$ in Eq.(\ref{eq:E3.12})
enters another identity that follows from the Dirac equation. It
arises after contracting indices $a$ and $b$ in
Eq.(\ref{eq:E3.7}),
\begin{eqnarray}\label{eq:E3.13}
D_c[\psi^+ \alpha^a {\overrightarrow D}_a \psi]\!=\!
\partial_c[\psi^+ \alpha^a {\overrightarrow D}_a \psi]\!- \omega_{abc}\psi^+
\alpha^b {\overrightarrow D}_a \psi.~~~
\end{eqnarray}
Eq.(\ref{eq:E3.13}) reveals one more inconsistency, which is
similar to the one observed in Eq.(\ref{eq:E3.8}). The quantity
$D_cT^a_{~a}$ is derivative of the trace of a tensor, i.e. of a
scalar. However the result has an additional term with a
connection, which is one more piece of evidence that the
quantities, $T^a_{~b}$, are not the invariants of a tensor. By the
same token, the l.h.s. of Eq.(\ref{eq:E3.13}) is not a covariant
derivative of a scalar.

By virtue of the Dirac equation, the first term on the r.h.s. of
(\ref{eq:E3.13}) becomes $\partial_c[-im\psi^+\rho_1\psi]$.
Alternatively, one can immediately use the equations of motion on
the l.h.s. and only then differentiate,
\begin{eqnarray}\label{eq:E3.14}
D_c[\psi^+ \alpha^a {\overrightarrow D}_a \psi]= -im
D_c[\psi^+\rho_1 \psi] \nonumber\\ = -im\partial_c[\psi^+\rho_1
\psi]+ im \psi^+ [\Gamma^+_c \rho_1+\rho_1\Gamma_c]\psi ~.
\end{eqnarray}
Comparing the last two equations and using (\ref{eq:A.4}) we
finally get
\begin{eqnarray}\label{eq:E3.15}
\omega_{acb}\cdot T_{ca} =2mg{\cal P}\aleph_b .
\end{eqnarray}
Using Eq.(\ref{eq:E3.15}), one can then rewrite
Eq.(\ref{eq:E3.12}) in a formally covariant form,
\begin{eqnarray}\label{eq:E3.16}
\nabla_\sigma T^\sigma_{~\nu}= i \psi^+\alpha^\mu[D_\mu, D_\nu]
\psi +2mg{\cal P}\aleph_\nu~, ~~
\end{eqnarray}
where $\aleph_\nu= e_\nu^{(a)}\aleph_a$ and the commutator $D_\mu
D_\nu-D_\nu D_\mu=[D_\mu, D_\nu]=-e^a_\mu e^b_\nu \mathbb{D}_{ab}$
on the r.h.s. has no derivatives. For the sake of completeness we
mention that the imaginary part  of $T^\sigma_{~\nu}$ is a tensor;
it is the covariant derivative $(i/2)\nabla_\nu j^\sigma$. Because
the vector current is conserved, the imaginary part of
Eq.(\ref{eq:E3.16}) is just an identity \cite{Fock1}, $
i\nabla_\sigma(\nabla_\nu j^\sigma)=i R_{\sigma\nu} j^\sigma$,
where $R_{\mu\sigma}$ is the Ricci curvature (contracted Riemann
tensor of curvature).

An attempt to make $\aleph_\nu =0$ leads to the main result of
Fock's paper \cite{Fock1}, which was derived entirely in the
coordinate representation (using $D_\mu$ as a well-defined
operator) and interpreted, with the reference to the
correspondence principle, as the equation of a geodesic line.
Since Eqs.(\ref{eq:E3.9}) and (\ref{eq:E3.15}) are nothing but two
identities that follow from the Dirac equation and
Eq.(\ref{eq:E3.16}) is their sum, there obviously is a way to
derive Eq.(\ref{eq:E3.16}) in one step, which was done by Fock
(with a subtle inaccuracy). The possibility to set $\aleph_a =0$
can be viewed as evidence that the abnormal term
$\omega_{cab}T_{ac}$ is zero. Such an impression is not correct.
This would require that $\omega_{cab}=0$. If the second term in
the r.h.s. of Eq.(\ref{eq:E3.12}) is zero so is the first one.
There remains nothing to move to the left in order to compile the
covariant derivative of the tensor.

Consider another possibility that the Riemannian spacetime, which
is hosting a Dirac field configuration (like proton) admits an
orthogonal system of coordinate hypersurfaces. Then the Ricci
coefficients with all different ordinal numbers vanish, i.e.
$\omega_{cab}-\omega_{cba}=0$, and the first two terms in the
r.h.s. of identity (\ref{eq:E3.12}) just cancel each other. Once
again, the possibility to compile the covariant derivative of
$T_{ab}$ as a part of the conservation law is lost. We are led to
conclusion that {\em the metric of spacetime, which is  hosting
the Dirac field (and thus is determined by this field) does not
allow for a system of orthogonal coordinate surfaces}.

If we formally translate the remaining terms into the coordinate
representation, we  get, instead of (\ref{eq:E3.12}) and
(\ref{eq:E3.15}),  two equations,
\begin{eqnarray}\label{eq:E3.17}
\partial_\sigma T^\sigma_\mu+\Gamma^\sigma_{\nu\sigma}T^\nu_{~\mu}-
(\partial_\sigma e_\mu^a)e_a^\nu
T^\sigma_{~\nu}~~~~~~~~~~~~~~~~~~~~~~ \nonumber\\ = i \psi^+
\alpha^\mu[D_\mu D_\nu- D_\nu D_\mu ]\psi,~~~~~~~~  \\
\label{eq:E3.18} (\nabla_\mu e_\sigma^a)e_a^\nu T^\sigma_{~\nu}= [
e^\nu_a\partial_\mu e^a_\sigma -\Gamma^\nu_{\sigma\mu}]
T^\sigma_{~\nu} =2mg{\cal P}\aleph_\mu,~
\end{eqnarray}
both carrying an explicit dependence on the tetrad vectors. The
sum of these equations reads as,
\begin{eqnarray}
\partial_\sigma
T^\sigma_\mu+\Gamma^\sigma_{\nu\sigma}T^\nu_{~\mu}
-\Gamma^\nu_{\sigma\mu} T^\sigma_{~\nu}- (\partial_\sigma
e_\mu^a-\partial_\mu  e^a_\sigma)e_a^\nu
T^\sigma_{~\nu}~\nonumber\\= i \psi^+ \alpha^\mu[D_\mu D_\nu-
D_\nu D_\mu ]\psi + 2mg{\cal P}\aleph_\mu, \nonumber
\end{eqnarray}
where this dependence is apparently hidden because we assumed that
all congruences are normal and $\partial_\sigma e_\mu^a-
\partial_\mu e^\sigma_a = e_\sigma^c(\omega_{abc}-
\omega_{acb})e^b_\mu=0$. This equation indeed coincides with
(\ref{eq:E3.16}), but only when the connection,
$\Gamma^\nu_{\sigma\mu}$, is symmetric, which was not an {\em a
priori} requirement. Since, in general, the Ricci coefficients are
not zero and the \textquotedblleft tensor" $T_{\sigma\mu}$ is not
symmetric (except for a plane-wave solution) , we cannot argue
that the r.h.s. of (\ref{eq:E3.15}) must be zero for whatever
reason. If, in addition to (\ref{eq:A.1}), we unconditionally
required that $\delta{\cal S}=\delta{\cal P}=0$, then arriving at
(\ref{eq:E3.15}) we would generate controversy.

An {\em ad hoc} choice of an {\em orthogonal coordinate system}
(where $\omega_{abc}=\omega_{acb}$) can serve only as a crude
approximation. To be consistent, we have to replace
$\omega_{acb}T_{ca}$ by $\omega_{abc}T_{ca}$ in
Eqs.(\ref{eq:E3.12}) and Eq.(\ref{eq:E3.15}) simultaneously. Then,
the latter can be rewritten as $\omega_{bca}\cdot T_{ca}=-2mg{\cal
P}\aleph_b$, so that the symmetric (and only the symmetric) part
of $T_{ba}$ matters. This transmutation indicates that we
implicitly employed the approximation of a material point when
internal deformations, that are bringing an object into a new
state of motion, are discarded. In this case, the unit vector
$e^\mu_{(0)}$ plays the role of the 4-velocity $u^\mu$ of a small
object as a whole. Since the first term in brackets in
Eq.(\ref{eq:E3.18}) can be dropped, we may write
\begin{eqnarray}\label{eq:E3.19}
{\partial\over\partial x^\mu}\big[\sqrt{-g}~{\sf Re}(T^\mu_{~\nu})
\big] = e \sqrt{-g}j^\mu F_{\mu\nu}~, \\ \label{eq:E3.20}
\Gamma^\nu_{\sigma\mu} ~{\sf Re}(T^\sigma_{~\nu}) =-2mg{\cal
P}\aleph_\mu ~,
\end{eqnarray}
which is a perfect expression for the energy-momentum conservation
complemented by the constraint (\ref{eq:E3.18}). If the equation
$\nabla_\sigma T^\sigma_{~\nu}=0$ is considered as a prototype for
the equation of a geodesic line (as it was conjectured in
\cite{Fock1}) then the term $\Gamma^\nu_{\mu\sigma}
T^\sigma_{~\nu}= \Gamma^\nu_{\sigma\mu} T^\sigma_{~\nu} $ in it is
connected with $\aleph_\mu$ through Eq.(\ref{eq:E3.20}). Depending
on the nature of the physical process, this term is responsible
either for the gravitational force or for the force of inertia.
These forces are real and one cannot set $\aleph_\mu$ or  ${\cal
P}$ to zero without losing them. An estimate of the coordinate
dependence of $\aleph_\mu$ yields Newton's approximation for the
metric tensor.  At large distances, we have $\aleph\propto 1/r^2$,
as it follows from Eq.(\ref{eq:E4.17}). A startling connection of
the field ${\cal P}$ with the localization of the Dirac field and
origin of its mass is discussed in Sec.\ref{sec:Sec4} .

The physical origin of these forces can be understood from another
perspective. The r.h.s. of Eq.(\ref{eq:E3.18}) can be rewritten as
$m(\partial_a {\cal S}-D_a {\cal S})$. The first term accounts
only for propagation of the wave form considered as an object. The
second term also accounts for the internal polarization of the
wave field. The difference between them is a force, which is due
to internal polarization. This fact motivates the view on
coordinates, as descendants of the polarization structure of the
Dirac field, which was proposed in Sec.\ref{ssec:Sec2A}. Its
dynamics are described by Eq.(\ref{eq:B.10}). An immediate
consequence of the existence of an internal dynamic in a localized
Dirac waveform is a view of pions as one of polarizations of the
Dirac field, which can be resolved, e.g., as a $2\gamma$-resonance
(see Appendix \ref{app:appB}).

\section{\normalsize Dirac field and congruences of curves.  \label{sec:Sec4}}

In this section, we closely follow the ideas of the intrinsic
geometry of Ricci and Levi-Civita as they are presented in the
monograph \cite{TLC}; the metric properties of the spacetime are
expressed in terms of rotations of the local coordinate pyramid.
The main subject of the analysis are Eqs.(\ref{eq:E3.15}) and
(\ref{eq:B.4}), which are the differential identities that follow
from the Dirac equations. Eq.(\ref{eq:E3.15}) is trivially
satisfied only for plane waves, i.e., when ${\cal P}=0$ and the
tensor $T_{ab}$ is symmetric. These solutions are employed in
scattering theory and they do not represent particles. In such a
context, equations like (\ref{eq:E3.15}) and (\ref{eq:B.4}) cannot
even be derived. Unlike the commonly known identities
(\ref{eq:E3.3}) and (\ref{eq:E3.4}), Eqs.(\ref{eq:E3.15}) and
(\ref{eq:B.4}) are not covariant in the sense that they explicitly
depend on congruences, which are the physical characteristics of
the Dirac field. It appears that these identities impose important
limitations on the properties of the metric, which is compatible
with the localized solutions of the Dirac equations. These
limitations are studied in the following section.

\subsection{Vector current and timelike congruence. \label{ssec:Sec4A}}

The Ricci coefficients are real-valued and skew-symmetric in the
first two indices. The tensor $T_{ab}$ is neither real nor
symmetric. The r.h.s. of Eq.(\ref{eq:E3.15}) is real. Therefore,
the imaginary part of Eq.(\ref{eq:E3.15}) is just $ {\rm Im}
(T_{ac}-T_{ca})=D_c(\psi^+ \alpha_a\psi)-D_a(\psi^+\alpha_c\psi)
=\nabla_c j_a-\nabla_a j_c=0$, and it should be considered
together with the equation $\nabla_a j^a=0$ of the vector current
conservation. Since $\nabla_a j_b$ are the invariants of a true
tensor, $\nabla_\mu j_\nu$, we have two tensor  equations,
\begin{eqnarray}\label{eq:E4.1}
\nabla_\mu j_\nu - \nabla_\nu j_\mu =0 ~.
\end{eqnarray}
and
\begin{eqnarray}\label{eq:E4.2}
\nabla_\mu j^\mu =0~.
\end{eqnarray}
The vector field $j^\mu(x)$ is globally timelike; its tangent unit
vector is $e^\mu_{(0)}(x)$, $j^\mu ={\cal R}~e^\mu_{(0)}$, where
$~{\cal R} =\sqrt{j^2}>0~$ is the invariant density of the Dirac
(spinor) matter. Therefore, Eq.(\ref{eq:E4.1}) becomes
\begin{eqnarray}\label{eq:E4.3}
\nabla_\mu e^{(0)}_\nu - \nabla_\nu e^{(0)}_\mu  +
e^{(0)}_\nu\partial_\mu\ln{\cal R}-
e^{(0)}_\mu\partial_\nu\ln{\cal R}=0.~~
\end{eqnarray}
Contracting this equation with $ e_{(a)}^\nu  e_{(b)}^\mu$,
$a,b=1,2,3$ and recalling Eqs.(\ref{eq:E2.19}) we find that
\begin{equation}\label{eq:E4.4}
\omega_{0ab}-\omega_{0ba} =0~, ~~~~a,b=1,2,3~
\end{equation}
which is a necessary and sufficient condition for the congruence
$e_{(0)}^\mu$ to be normal \cite{TLC,Eisenhart}. Namely, there
exists such a function, ${\cal T}(x)$, that the vector field
$e^{(0)}_\mu (x)$ is orthogonal to the family of surfaces ${\cal
T}(x)={\rm const}$,
\begin{equation}\label{eq:E4.5}
\partial_\mu {\cal T}(x)= f(x)e^{(0)}_\mu (x),
\end{equation}
where $f(x)$ is a coordinate scalar. Contracting
Eq.(\ref{eq:E4.3}) with $ e_{(0)}^\nu$ we get
\begin{equation}\label{eq:E4.6}
\partial_\mu\ln{\cal R}=e^{(0)}_\mu\partial_{(0)}\ln{\cal R}-
\omega_{b00}e^{(b)}_\mu,
\end{equation}
where $\partial_{(0)}\ln{\cal R}=e_{(0)}^\mu\partial_\mu\ln{\cal
R}$, is the derivative in the direction of the arc $s_0$.
Contraction of Eq.(\ref{eq:E4.3}) with $ e_{(0)}^\nu e_{(a)}^\mu$
yields
\begin{equation}\label{eq:E4.7}
(\partial\ln{\cal R}/ \partial s_a)=-\omega_{a00}~,~~~a=1,2,3,
\end{equation}
which indicates that congruences of lines, defined by the system
of equations (\ref{eq:E2.14}),  $dx^\mu/ds_0=e_{(0)}^\mu ~$, must
experience permanent bending (acceleration) whenever the invariant
density ${\cal R}(x)$ of the Dirac field is not uniformly
distributed. The spatial gradient of ${\cal R}(x)$ cannot vanish
for any localized state. Even more, the congruence of lines of the
Dirac current is not a geodesic congruence, since, for geodesic
lines, the vector of geodesic curvature would have vanished, i.e.,
$\omega_{a00}=0$.

Additional information can be extracted from Eq.(\ref{eq:E4.2}).
From definition (\ref{eq:E2.20}) it follows that
\begin{equation}\label{eq:E4.8}
\nabla_\nu e_{(0)}^\nu = - (\partial\ln{\cal R}/ \partial s_0)
=\sum_a\eta_{(a)} \omega_{0aa}~.
\end{equation}
Hence, we can rewrite  (\ref{eq:E4.6}) as
\begin{equation}\label{eq:E4.9}
\partial_\mu\ln{\cal R}=-e^{(0)}_\mu\sum_a\eta_{(a)} \omega_{0aa}
-\omega_{b00}e^{(b)}_\mu,
\end{equation}
which shows that  the r.h.s. of Eq.(\ref{eq:E4.9}), which contains
only geometric objects, is a component of a gradient. Together
with condition (\ref{eq:E4.4}) this constitutes a necessary and
sufficient condition that the function  ${\cal T}(x)$, defined by
Eq.(\ref{eq:E4.5}), is an harmonic function \cite{Eisenhart},
\begin{equation}\label{eq:E4.10}
\Box~ {\cal T}=g^{\mu\nu}\nabla_\mu\nabla_\nu {\cal T}=0.
\end{equation}
We may take the parameter $t$ of ${\cal T}(x)=t=const$ as a
definition of the world time. For the harmonic function, ${\cal
T}(x)$, the conditions of integrability for the system
(\ref{eq:E4.5}) of partial differential equations reads as
\cite{Eisenhart} $$\partial_\mu\ln f =-e^{(0)}_\mu\sum_a
\eta_{(a)}\omega_{0aa} -\omega_{b00}e^{(b)}_\mu~.$$ Comparing it
with (\ref{eq:E4.9}) we find that $ f(x)={\cal R}$, so that the
world time $t$ and the \textquotedblleft proper time" $s_0$ are
related by
\begin{equation}\label{eq:E4.11}
 dt= {\cal R}ds_0 = ds_0/\sqrt{g_{00}}~.
\end{equation}
Hence, we can draw the major conclusion that: {\em The proper
time, $s_0$, flows more slowly than the world time, $t$, whenever
localized Dirac matter is present.} If Dirac matter is in a stable
configuration (on mass shell), then there is a well defined time
and one can consistently speak of a (quantum) state of the Dirac
field.

Since the congruence $e^\mu_{(0)}$ appeared to be normal, the
hypersurfaces ${\cal T}(x)=t=const$ represent space at different
times $t$. The three other vectors $e_{(i)}^\mu (x)$, $i=1,2,3$ of
local tetrad are spacelike, orthogonal to $e_{(0)}^\mu (x)$ and
thus belong to such hypersurfaces (by the definition,
$e_{(i)}^\mu\partial_\mu{\cal T}=0$). The interval becomes as
\begin{equation}\label{eq:E4.11a}
ds^2=g_{00}dt^2 + {\sf g}_{ik}dx^idx^k.
\end{equation}
Accordingly, $e_{(0)}^\mu =(1/\sqrt{g_{00}},\vec{0})$,
$e^{(0)}_\mu =(\sqrt{g_{00}},\vec{0})$.

Equation (\ref{eq:E3.3}) of the vector current conservation now
reads as
\begin{equation}\label{eq:E4.12}
\partial_\mu(\sqrt{-g}e_{(0)}^\mu {\cal R})=
\partial_t(\sqrt{-g}~g^{00})= 0.
\end{equation}
This can be recognized as the condition for the coordinate $x^\mu$
to be harmonic, $$ \Box \varphi={1\over\sqrt{-g}}
{\partial\over\partial x^ \nu} \bigg(\sqrt{-g}~g^{\mu\nu}
{\partial \varphi\over\partial x^ \mu}\bigg)=0,$$ which is
specified for the normal coordinate $\varphi={\cal T}=x^0$ (see,
e.g. \cite{Fock2}, \S 41). From (\ref{eq:E4.12}), there follows
one more (very intuitive) form of the current conservation,
\begin{equation}\label{eq:E4.13}
\partial_t({\cal R}\sqrt{-{\sf g}})=0,
\end{equation}
where ${\sf g}={\sf det}|{\sf g}_{ik}|$ is the determinant of the
spatial metric. This form means that when the density $\cal R$
grows and local time slows down, then the measure of space volume
shrinks. Since the variation of ${\cal R}$ and time slowdown both
are intimately connected with acceleration, the last equation
unites them and the Lorentz contraction of a localized object {\em
in one physical process}. One should not even refer to a spacelike
interval between two events on the opposite sides of an elementary
object.

\subsection{Axial current and spatial part of metric. \label{ssec:Sec4B}}

The axial current ${\cal J}^\mu$ is spacelike and orthogonal to
the vector $j^\mu$. According to Eq.(\ref{eq:E3.4}), the axial
current has a source $2m {\cal P}$, which is localized together
with the invariant density ${\cal R}$. Since there is no flux of
vector current in this direction (the amount of matter inside a
closed surface remains the same), we associate the radial
direction $e_{(3)}^\mu (x)$ with the axial current, ${\cal J}^\mu
={\cal R} e_{(3)}^\mu$. Then Eq.(\ref{eq:E3.4}) takes form
\begin{equation}\label{eq:E4.14}
\nabla_\mu e_{(3)}^\mu +  e_{(3)}^\mu \partial_\mu \ln{\cal R} =
2m {\cal P}/ {\cal R}= 2m \sin\Upsilon~.
\end{equation}
On the one hand, by definition, $$\nabla_\mu e_{(3)}^\mu=\sum_a
\eta_{(a)} \omega_{3aa}= \omega_{300}- \omega_{311} -
\omega_{322}.$$ On the other hand, according to
Eq.(\ref{eq:E4.7}), we have $$e_{(3)}^\mu \partial_\mu \ln{\cal
R}=\partial \ln{\cal R}/\partial s_3=-\omega_{300}.$$ Substituting
these expressions into Eq.(\ref{eq:E4.14}) we obtain an extremely
important relation,
\begin{equation}\label{eq:E4.15}
\omega_{131} + \omega_{232} =2m \sin\Upsilon~.
\end{equation}
This can be read in different ways. First and foremost, it
expresses the curvature of the two-dimensional surface $(s_1,s_2)$
of angular coordinates via the local parameter $\sin\Upsilon={\cal
P}/{\cal R}$ of the Dirac field. {\em Vice versa}, once geodesic
curvatures $\omega_{131}$ and $\omega_{232}$ are known in advance
(e.g., from an alleged symmetry, experiment, etc.) then
$\Upsilon(x)$ is known as an explicit function of spacetime
coordinates and there remains no freedom of  \textquotedblleft
chiral" transformations, like in Eq.(\ref{eq:A.8}).

Second, the l.h.s. of (\ref{eq:E4.15}) must be a well-defined
geometric object (at least when the congruence $e_{(3)}^\mu$ is
normal and the radial coordinate  is well-defined). In this case,
we must have $D_a\Upsilon=0$ because the covariant differential
operator $D_a$ is defined only by its action on the Dirac field.
Consequently, by virtue of Eq.(\ref{eq:A.8}), $D_a\Upsilon[\psi]=
\partial_a\Upsilon + 2g\aleph_a$, we have
\begin{eqnarray}\label{eq:E4.16}
2g\aleph_a=-\partial_a\Upsilon,
\end{eqnarray}
i.e. the field $\aleph_a$ must be a gradient. If the congruence
$e_{(3)}^\mu$ is not normal, then any symmetry of any explicit
solutions of the Dirac equation with respect to arcs $s_1$ and
$s_2$ should be considered {\em a dynamical internal symmetry}.

Third, for a concave surface the curvature is positive so that
$0<\Upsilon<\pi/2$. For the normal orthogonal spherical
coordinates we have $\omega_{131} = \omega_{232}=1/r$ and if such
a coordinate system were possible we would immediately know that
\begin{eqnarray}\label{eq:E4.17}
\Upsilon[\psi]=\arcsin(1/mr),~~~mr>1\nonumber\\2g\aleph_3[\psi] =-
\partial_r \Upsilon={1\over r\sqrt{m^2r^2-1}}.
\end{eqnarray}
Obviously, this simple formula cannot be exact; rather it predicts
the correct asymptotic behavior at large distances.

 Fourth, the condition $\sin\Upsilon(x)<1$ defines the mass parameter
$m$ as the upper limit of a possible curvature, which is, in fact,
a {\em definition of mass from the perspective of the internal
structure of a Dirac particle}. (In spherical case we would have
$mr>1$; the radius must exceed the Compton length!) This result is
also in agreement with the {\em kinematic} Lorentz contraction of
special relativity. An accelerated particle is Lorentz contracted
and both ${\cal R}$ and the maximal curvature become $\propto
1/\sqrt{1-v^2}$.

In order to facilitate further analysis of the real part of
Eq.(\ref{eq:E3.15}), let us rewrite it's l.h.s in terms of the
axial current. Let us use the dual representation of the axial
current as $\epsilon^{stua}{\cal J}_a= i\psi^+\alpha^s\rho_1
\alpha^t\rho_1 \alpha^u \psi$, $~(s,t,u,\neq)$, and differentiate
it. The result reads as
\begin{eqnarray}\label{eq:E4.18a}
D_u \epsilon^{stua}{\cal J}_a = i \sum_{u\neq s,t} D_u
(\psi^+\alpha^s\rho_1 \alpha^t\rho_1 \alpha^u \psi).
\end{eqnarray}
If we extend here the sum over all values of $u$ (this sum
vanishes by virtue of the equations of motion) and subtract the
terms with $u=s$ and $u=t$, the result will be
\begin{eqnarray}
D_u \epsilon^{stua}{\cal J}_a = -i \psi^+\alpha^s {\overrightarrow
D}_t\psi + i\psi^+\alpha^t {\overrightarrow D}_s\psi \nonumber \\
-i \psi^+{\overleftarrow D}^+_s\alpha^t\psi +i
\psi^+{\overleftarrow D}^+_t\alpha^s\psi~,\nonumber
\end{eqnarray}
where the r.h.s is four times the anti-symmetric Hermitian part of
the energy momentum tensor. Therefore, the real part of
Eq.(\ref{eq:E3.15}) reads as
\begin{equation}\label{eq:E4.18}
(1/4) \omega_{acb}\cdot \epsilon^{acst}\cdot\nabla_s{\cal J}_t
=2mg{\cal P}\aleph_b,
\end{equation}
and can be immediately recognized as the dual to
Eq.(\ref{eq:B.5}), derived for the stress tensor,
\begin{eqnarray}
~~~~~~~~~~~~~~~~(1/2)\omega_{acb} \nabla_c{\cal J}_a =-2g m {\cal
S}\aleph_b.~~~~~~~~~~~~{\rm (B.5)}\nonumber
\end{eqnarray}

These two equations clearly indicate that {\em any motion of the
Dirac field follows the path of a helix}. The underlying reason
for such a motion is the non-vanishing curl of the spacelike axial
current. The boost $\omega_{0ia}$ in the direction $s_i$ is
inevitably accompanied by the spatial rotation
$\epsilon^{ijk}\omega_{jka}$ in the plane perpendicular to $s_i$.
In plain words, {\em the Dirac field cannot be accelerated without
causing a rotation thus behaving as a (relativistic) system of
inertial navigation}.

In Eqs.(\ref{eq:E4.18}) and  (\ref{eq:B.5}), $\nabla_s{\cal J}_t=
e_{(s)}^\mu e_{(t)}^\nu\nabla_\mu {\cal J}_\nu$. Since ${\cal
J}^\mu= {\cal R} e_{(3)}^\mu $, we further have
\begin{eqnarray}
D_s{\cal J}_t={\cal R}e_{(s)}^\mu e_{(t)}^\nu\nabla_\mu
e^{(3)}_\nu + \delta^3_t e_{(s)}^\mu\partial_\mu {\cal R}
\nonumber\\ =  {\cal R} [\omega_{3ts} + \delta^3_t (\partial
\ln{\cal R}/\partial s_s)]~.~~~\nonumber
\end{eqnarray}
Finally, using  Eqs.(\ref{eq:E4.7}) and (\ref{eq:E4.8}), which
define $\omega_{s00}$ and $\omega_{0aa}$ as the functionals of the
Dirac field, we arrive at
\begin{widetext}
\begin{eqnarray}\label{eq:E4.19}
(1/4)~\omega_{acb}\cdot \epsilon^{acst}[\omega_{3ts} - \delta^3_t
\omega_{s00}- \delta^3_t\delta^0_s\omega_{0aa}]=
m\sin\Upsilon\cdot 2g \aleph_b,\\\label{eq:E4.20}
(1/2)~\omega_{stb}\cdot[\omega_{3ts} - \delta^3_t \omega_{s00}-
\delta^3_t\delta^0_s\omega_{0aa}]= -m\cos\Upsilon\cdot 2g
\aleph_b,
\end{eqnarray}
\end{widetext}
At this point, we can conclude that the parameters $\aleph_b$ in
the connection $\Gamma_b$ (\ref{eq:A.3}) are totally defined by
the bending of the system of congruences. Despite that fact that
the Lagrangian for the Dirac equations (\ref{eq:E3.1}) and
(\ref{eq:E3.2}) includes the term ${\cal J}^a \aleph_a$, which can
be interpreted as an interaction between the axial current and the
field $\aleph_a$, it cannot be viewed as an independent field that
is governed by an additional equation of motion. (Otherwise, such
equations must be {\em invented}, which we, so far, tried to
avoid.)

\subsection{The case of the normal radial coordinate.
Qualitative consequences of the localization. \label{ssec:Sec4C}}

Even for the localized waveforms, the existence of the surface of
a constant distance from a center is not given {\em gratis}. Such
a surface must be orthogonal, at every point, to the tangent
vectors $e^\mu_{(3)}$ of the congruence of lines of the axial
current. Unlike the previously studied case of the timelike
congruences  $e^\mu_{(0)}$, the corresponding conditions for
integrability do not universally follow from the equations of
motion. Most likely, the radial coordinate cannot be normal.
However, sometimes (mostly for long-lived particles) empirical
data may hint that such a normal hypersurface of a constant
distance $s_3$ from a center, which is spanned by the
\textquotedblleft angular" arcs $(s_1,s_2)$ with tangent unit
vectors (\ref{eq:E2.7}) may be a good approximation. This is what
we intuitively expect in a one-body problem and we have to verify
that this assumption is consistent with the equations of motion
and the established earlier constraints. In what follows, we
consider Eqs.(\ref{eq:E4.17}) as the criterion of the spherical
symmetry and try to profit from the fact that Eqs.(\ref{eq:E4.19})
and (\ref{eq:E4.20}) significantly simplify under the assumption
that the congruence $e_{(3)}^\mu$ is a normal congruence (which,
possibly, can be a first approximation in a sequence of
iterations). This will also allow us to qualitatively understand
the trends in critical behavior of the tetrad vectors near the
limit surface, $\sin\Upsilon\to 1$, and rediscover some well known
properties of matter (which cannot be done in a picture of matter
as plane waves).

Since the congruence $e_{(3)}^\mu$ is set normal, there should
exist  a function ${\cal L}(x)$ such that
\begin{equation}\label{eq:E4.21}
\partial_\mu {\cal L}(x)= \zeta(x)e^{(3)}_\mu (x),
\end{equation}
where $\zeta(x)$ is a coordinate scalar.  The hypersurfaces $
{\cal L}(x)=r=const$ are the surfaces of radius $r$, i.e.,
\begin{equation}\label{eq:E4.22}
dr= \zeta ds_3 = ds_3/\sqrt{g_{33}}~.
\end{equation}
From the integrability condition for Eq.(\ref{eq:E4.21}) it is
straightforward to derive the equations (which are similar to
equations for the the function $f(x)$ (cf.(\ref{eq:E4.5}))
\begin{equation}\label{eq:E4.23}
-\partial_\mu\ln\zeta=e^{(3)}_\mu\partial_{(3)}\ln\zeta-
\omega_{a33}e^{(a)}_\mu~,
\end{equation}
\begin{equation}\label{eq:E4.24}
(\partial\ln\zeta/ \partial s_a)=\omega_{a33}~,~~~(a=0,1,2) ;
\end{equation}
but, we have no constraint that would express $\zeta$ as a
function of the Dirac field. For the normal congruence
$e_{(3)}^\mu$ we  have $\omega_{3ab}= \omega_{3ba}$ for $a,b\neq
3,~~a\neq b$, as a necessary and sufficient condition
\cite{TLC,Eisenhart} and, consequently, the first term in brackets
in Eqs.(\ref{eq:E4.19}) and (\ref{eq:E4.20}) simplifies,
$\omega_{stb}\omega_{3st}=-\omega_{03b}\omega_{033}+\omega_{13b}\omega_{133}
+\omega_{23b}\omega_{233} $ and $\omega_{acb}\epsilon^{acst}
\omega_{3st} =\omega_{12b}\omega_{033}+\omega_{02b}\omega_{133}
+\omega_{01b}\omega_{233}$. In the second term, the sum includes
only $a,c=1,2$. Because we assume a stable object, the third term
in brackets cancels, $\partial\ln({\cal R}\zeta)/
\partial s_0\approx 0$. Hence, the
system of Eqs.(\ref{eq:E4.19}) and (\ref{eq:E4.20}) can be cast as
\begin{eqnarray}\label{eq:E4.25}
(\omega_{1}\omega_{02b}-\omega_{2}\omega_{01b}) = 2mg\aleph_b
\sin\Upsilon , \nonumber \\
(\omega_{1}\omega_{31b}+\omega_{2}\omega_{32b}) =
2mg\aleph_b\cos\Upsilon,
\end{eqnarray}
where $$\omega_j={\omega_{j00}+\omega_{j33}\over 2}={1\over 2}
{\partial\ln(\zeta /{\cal R})\over\partial s_j},~~ j=1,2.$$ Then,
Eqs.(\ref{eq:E4.25}) with $b=0,1,2$ (and $\aleph_0\!=\aleph_1
\!\!=\aleph_2=0$) yield a set of six equations,
\begin{eqnarray}\label{eq:E4.26}
{\omega_{1}\over \omega_{2}}=-{\omega_{320}\over \omega_{310}}
={\omega_{011}\over \omega_{021}} ={\omega_{012}\over
\omega_{022}}={\omega_{321}\over \omega_{131}} ={\omega_{232}\over
\omega_{312}}=\pm 1,\nonumber\\
\end{eqnarray}
where the rightmost equation  immediately follows from
$\omega_{312}^2=\omega_{131}\omega_{232}=1/r^2$ and
$\omega_{131}=\omega_{232}=1/r$. In spherical case, $\aleph_3$ is
given by Eqs.(\ref{eq:E4.17}) and, consequently,
\begin{eqnarray}\label{eq:E4.27}
2\omega_{1}\omega_{023} = {1\over r^2\sqrt{m^2r^2-1}}~,~~
\omega_{1}(\omega_{313}\pm\omega_{323}) = {1\over r^2},\nonumber
\\ \omega_{2}=\pm\omega_{1},
~\omega_{013}=\mp\omega_{023},~\omega_{312}=\omega_{321}=\pm 1/r
.~~~~~
\end{eqnarray}
As one could expect, in the case of normal radial congruences,
there is a full symmetry between congruences of arcs $ds_1$ and
$ds_2$.
At large distances we generally have $mr\gg 1$ so that spatial
rotations dominate. {\em Vice versa}, near the inner boundary
$mr=1$ the boosts $\omega_{013}ds_3$ and $\omega_{023}ds_3$  in
tangent directions (as well as boosts $\omega_{031}ds_1$ and
$\omega_{032}ds_2$  in radial directions) become infinite. When,
starting from a generic point $x_0$, we move along lines of
congruences $e_{(i)}^\mu(x)$ approaching $r\sim m^{-1}$, then
local tetrad  rotate (with respect to the $e_{(i)}^\mu(x_0)$) in
such a way that {\em all} directions become nearly lightlike, so
that the tangent velocities $v_i=\dot{s}_i\to c$. These
observations explain the formally derived inner boundary $r=1/m$
(generally, $\sin\Upsilon=1$) of the Dirac particle as the caustic
of the lines of the Dirac currents.

In fact, we have two interconnected mechanisms of the time
slowdown (due to the amplified ${\cal R}$ and because the vector
currents tend to approach the lightlike directions), which cannot
be separated. From the perspective of an \textquotedblleft
external observer", the time flow literally stops at the critical
surface of a stable Dirac waveform. Therefore, the sharp
interaction of the deeply inelastic scattering always resolves an
apparently static object in a random configuration determined by
the foregoing causal evolution (cf. discussion of evolution
equations of QCD in Ref.\cite{MS1}). Certain patterns of symmetry
observed in such processes most likely correspond to the symmetry
of (the critical points of) projection of the actual currents onto
a surface determined by the collision axis, the collision plane,
etc. These patterns well may have very little to do with the
internal dynamics of the stable waveform.

When $mr\gg 1$, we generally have $\sin\Upsilon\to 0$ and,
according to Eq.(\ref{eq:E2.10}), the pseudoscalar density nearly
vanishes and ${\cal S}\approx{\cal R}$. Magnetic polarization of
the Dirac field is  greater than the electric one, $|{\vec L}|\agt
|{\vec K}|$. {\em Vice versa}, at the shortest possible distances,
when $\Upsilon\to \pi/2$, the Lorentz boosts play a major role.
Accordingly, the pseudoscalar density ${\cal P}\approx{\cal R}$ is
large  and electric polarization is dominant, ${\vec K}^2-{\vec
L}^2\approx {\cal R}^2\gg 1$. These two sectors are separated by
the two-dimensional surface $\Upsilon[\psi]= \pi/4$. When
$\Upsilon[\psi]< \pi/4$ (large distances) an appropriate choice
for $e^\mu_{(1)}$ and $e^\mu_{(2)}$ will be vectors $H_i$ and
${\overstar H}_i$ of Eqs.(\ref{eq:E2.7}).  At the interval between
$\lambda$ and $\sqrt{2}\lambda$, where the mass of the particle is
being formed, these will be $E_i$ and ${\overstar E}_i$, defined
by a system of boosts, which are large along with the pseudoscalar
density.

In fact, this domain is also responsible for the magnetic moment
of the Dirac particle. Indeed, using the classical formula for the
magnetic moment, $\mu=(e/2c)[r\times v]$, with the limit values,
$r=\lambda=\hbar/mc$ and $v=c$, which follow from the
Eqs.(\ref{eq:E4.27}), we obtain
\begin{equation}\label{eq:E4.28}
\mu= {e\over 2c}\cdot {\hbar\over mc} \cdot c ={e\hbar\over 2mc} .
\end{equation}
This is a well-known result corresponding to the gyromagnetic
ratio $g=2$.

The impossibility to introduce a normal orthogonal coordinate
system in the presence of the axial potential $\aleph$ in the
equations of motion is explicitly illustrated in Appendix
\ref{app:appC}. An attempt to separate the angular variables in
the Dirac equation in the presence of only the radial component
$\aleph_r(r)$ is made. In this case, the radial coordinate, $r$,
is a well defined normal coordinate. It appears that even in such
a simplest case there are no operators with eigenvalues of angular
momentum that commute with the Hamiltonian. At the same time,
angular variables can be explicitly separated in the equations of
motion. The only possible explanation of this fact is that the
formally introduced angles do not represent arcs of the usual
spatial angular coordinates. There is no reason to require that
solutions of the Dirac equations must be single-valued along these
arcs. However, equations (\ref{eq:C.8}) for angular functions are
clearly the equations for spherical harmonics. These harmonics can
be interpreted only as elements of an internal symmetry in the
space of polarizations of the Dirac field. In order to find out
what can be the  non-geometric integrals of motion the angular and
radial functions must be studied together.

\section{\normalsize Nonlinear Dirac equation. \label{sec:Sec5}}

In this section, following the programme outlined in
Sec.\ref{ssec:Sec2B}, we will incorporate the nonlinear effects,
which so far were found as constraints, into the Dirac equation.
Following Fock \cite{Fock1}, let us rewrite the operator $\alpha^a
\Omega_a$ as
\begin{equation}\label{eq:E5.1}
\alpha^b \Omega_b = (1/2) \omega_{aca}\alpha^c - (i/4)
\epsilon_{acbd} \omega_{acb} \rho_3 \alpha^d.
\end{equation}
Then, the Dirac equation reads as
\begin{equation}\label{eq:E5.2}
\alpha^b \bigg[ {\partial \over \partial s_b} + ieA_b +ig
\rho_3\aleph_b - {1\over 2} \omega_{aba} - {i\over 2} {\bm w}_b
\rho_3\bigg]\psi =-im\rho_1\psi,
\end{equation}
where  $\aleph_b$ and ${\bm w}_b=-(1/2)\epsilon_{acdb}
\omega_{acd}$ are the sets of invariants; the latter differ from
zero whenever spacetime does not admit a coordinate net of all
normal congruences. In general, the invariants $w_b$ do not vanish
and they are complementary to the invariants $\aleph_b$ given by
Eqs. (\ref{eq:E4.19})and (\ref{eq:E4.20}). [Since congruence
$e_{(0)}^\mu$ is normal, we have ${\bm w}_1=-\omega_{230}$, ${\bm
w}_2= -\omega_{310}$, ${\bm w}_3=-\omega_{120}$.]

According to Eqs.(\ref{eq:E4.7}) and (\ref{eq:E4.8}), each sum
$\sum_a\eta_{(a)}\omega_{aba}$ includes either the terms
$\omega_{0i0}=\partial\ln{\cal R}/\partial s_i$ ($i=1,2,3$) or
$\sum_i\omega_{i0i}=\partial\ln{\cal R}/\partial s_0 $, all of
which are bilinear forms of $\psi^+$ and $\psi$ and,
geometrically, are the geodesic curvatures. At first glance, the
presence of these curvatures makes the Dirac equation extremely
non-linear. This genuine nonlinearity, however,  can be
effectively alleviated after the following \textquotedblleft
normalization"  of the Dirac wave function. If we observe that
\begin{eqnarray}\label{eq:E5.4}
{\partial \psi \over \partial s_a} - {1\over 2}{\partial \ln{\cal
R}\over \partial s_a}\psi =\sqrt{\cal R} {\partial \over \partial
s_a}\bigg({\psi\over \sqrt{\cal R}}\bigg)~
\end{eqnarray}
and assume Eq.(\ref{eq:E4.16}) be true, then we arrive at a much
simpler equation for the normalized function $\xi=\psi/ \sqrt{\cal
R}= (g_{00}[\psi])^{1/4}\psi$ :
\begin{widetext}
\begin{eqnarray}\label{eq:E5.5}
\bigg[i {\partial \over \partial s_0} -eA_0 -{1\over 2}\rho_3
[{\bm w}_0 +\partial_0\Upsilon]-
\sum_{i=1}^3\alpha^i  \bigg( i {\partial \over \partial s_i}
+ i {k_i\over 2} - eA_i 
+{1\over 2}\rho_3[{\bm w}_i+\partial_i\Upsilon]
\bigg)-m\rho_1\bigg]\xi=0,
\end{eqnarray}
where $k_i=\sum_{j\neq i}\omega_{jij}$. The nonlinear  Dirac
equation (\ref{eq:E5.2}) now looks like a {\em linear equation}
for the normalized function $\xi$. Once $m^{-1}$ is accepted as a
measure of length, this equation is dimensionless and does not
change under a similarity transformation. At the first glance, the
term $\rho_3\partial_i \Upsilon$ in it is always nonlinear;
however, the constraint $k_3=2m\sin\Upsilon=2m (\xi^+\rho_2\xi)$
(cf. Eq.(\ref{eq:E4.15})) eliminates the non-linearity whenever
the curvature $k_3$ can be determined as a function of coordinates
from geometric considerations. In the one body problem this
curvature always has the meaning of the inverse radius of the
enveloping convex surface. At least at large distances (at the
scale of $1/m$) we have $\Upsilon\propto 1/mr$ and
$\partial_r\Upsilon\propto -1/mr^2$, which brings in a singular
potential $\propto 1/r^2$ into the Dirac equation and a Newton's
force into Eqs.(\ref{eq:E3.15})-(\ref{eq:E3.18}). Such a construct
is equivalent to the Newton's approximation in general relativity
and it may serve as the first step of an iterative procedure for
the Dirac equation.

In general, there is no direct connection between the radius of
curvature and the distance to any distinct point inside a
localized object; the gradients $\partial_i \Upsilon$ can be
arbitrary large. Most likely, the solutions of Eq. (\ref{eq:E5.5})
have multiple caustics where the invariant density ${\cal R}$ is
large and ${\cal P}$ dominates to the extent that ${\cal
R}\approx{\cal P}$. So far, we did not find in mathematical
literature regular methods to study equations like
(\ref{eq:E5.5}). The methods of contact geometry \cite{Arnold}
seem to be most relevant.

The most important source of the nonlinearity of the Dirac
equation resides in that fact that evolution, in terms of proper
time $s_0$, has a different rate at different points of the
localized object; the rate of evolution, $\partial/\partial s_0$,
along {\em this} time cannot yield anything like the energy of
this object as a whole. Fortunately, we have proved that a stable
object does have a well defined hypersurface of a constant world
time $t$. Therefore, a meaningful evolution scale for a stable
object as a whole is associated with the physical time $t$.
According to Eq.(\ref{eq:E4.11}), we have $dt={\cal R} ds_0$.
Hence,
\begin{eqnarray}\label{eq:E5.6}
\bigg[i {\cal R} {\partial \over \partial t} -eA_0
-{1\over 2}\rho_3 [{\bm w}_0 +\partial_0\Upsilon] 
- \sum_{i=1}^3\alpha^i  \bigg( i {\partial \over \partial s_i}
+ i {k_i\over 2} - eA_i 
+{1\over 2}\rho_3 [{\bm w}_i+\partial_i\Upsilon]
\bigg)-m\rho_1\bigg]{\psi\over\sqrt{\cal R}}=0,
\end{eqnarray}
\end{widetext}
and now this equation has a scale fixing factor ${\cal R}$ in
front of the $\partial/\partial t$. This effect is a major one
because it corresponds to a clearly understood physical effect,
refraction of the Dirac waves. It is of the same physical origin
as self-focusing in nonlinear optics and acoustics or deflection
of light in the gravitational field of a star. In fact, ${\cal R}$
plays a role of  \textquotedblleft refractive index" depending on
the amplitude. Since the phase velocity decreases with increasing
amplitude, the field tends to auto-localize. This mechanism of
concentration is the most distinctive property of gravity which
may signal its role in matter formation from fields at all scales.
The spatially uniform solutions of the Dirac equation just cannot
be stable.

\section{\normalsize Localized Dirac matter,
 electric charge and Radiation. \label{sec:Sec6}}

Our major perception regarding the vacuum is the absence of
matter. Since matter inevitably is localized, this means that in
the vacuum, ${\cal R}$ is constant and the spacetime metric of the
Lorentz vacuum has $g_{00}\!=\!c^2\!=\!1$. At the same time, we
have Eq.(\ref{eq:E4.11}), $$~~~~~~~~~~~~~~~~~ dt=  {\cal R}ds_0 =
ds_0/\sqrt{g_{00}}~.~~~~~~~~~~~~~~(\ref{eq:E4.11})$$ Therefore,
the empirical $g_{00}=1$ of an empty space corresponds to ${\cal
R}=1$. This state cannot be stable. Due to a very special
nonlinearity of Eq.(\ref{eq:E5.6}) Dirac waves tend to refract
towards domains where ${\cal R}-1>0$ amplifying ${\cal R}$ there
until some saturation level (or caustic) is reached and an
external boundary is formed. The opposite trend must be observed
in domains where ${\cal R}-1<0$; the Dirac waves tend to escape
them. This conjecture can be phrased more precisely as:\\ {\em
Identification of the sign of $({\cal R}-1)$ with the sign of
electric charge leads to a dynamic picture of an empirically known
charge-asymmetric world in which stable positively charged
on-mass-shell Dirac objects are highly localized (and presumably
heavy) while negatively charged objects tend to be poorly
localized (and presumably light).} \\ The best prospect of this
idea is that these objects are the protons (or nuclei) and the
electrons of the real world. When electric forces come into play,
the electrons become somewhat localized around heavy objects, thus
forming electrically neutral matter. For atomic electrons, which
are smeared over distances much exceeding the Compton length and
held weakly localized near nuclei by electric forces, the effect
of ${\cal R}-1\ll 1$ on the metric must be negligible. However,
the view of a vacuum as a classical Dirac field with the standard
level ${\cal R}=1$ and propagating in it waveforms, instead of an
ensemble of quantum oscillators with an unbound spectrum that
interact with plane waves, may alleviate the problem of
ultraviolet divergences of the standard perturbation series and
its renormalization.

Our major conclusion about the nature of localization is drawn
from the analysis of the on-mass-shell waveforms. However, it
relies on basic properties of the wave propagation so that it
seems reasonable to apply them to at least the long-lived
waveforms. From this perspective, any positively charged particle
should have a slightly longer lifetime and be more localized than
its negative partner. While the proton is small and stable, the
anti-proton should not have an as well defined outward boundary as
the proton has. The lifetime of the anti-hydrogen may not be long
even when it is completely isolated from normal matter. We refrain
here from speculating about the possible mode of its decay.

The next important question is the interaction of the localized
Dirac waveforms with the electromagnetic field. Empirically, the
field $A_c$ in the connection (\ref{eq:A.3}) is the
electromagnetic field, which is responsible for the Lorentz force.
The system of invariants $\aleph_a$ in $\Gamma_a$ is determined
either by geometric properties of congruences like
(\ref{eq:E4.15}) or by nonlinear constraints (\ref{eq:E4.19}) and
(\ref{eq:E4.20}). As long as the field $\aleph_a$ is a gradient,
the r.h.s. of Eq.(\ref{eq:E3.9}) is a system of invariants, which
include, except for the geometric terms, the term $ej^aF_{ab}$. In
the coordinate representation (\ref{eq:E3.14}), this part is
translated into $ej^\mu F_{\mu\nu}$, and this is the only term on
the r.h.s. of the {\em real part} of Eq.(\ref{eq:E3.9}). This term
is known as the Lorentz force of the field $F_{\mu\nu}=\nabla_\mu
A_\nu-\nabla_\nu A_\mu$ acting on a charged particle. The vector
field $A_\mu=e^{(a)}_\mu A_a$ originates from the connection
$\Gamma_a$ and thus is an external field.

For the real part of the energy-momentum tensor, $T_{ab}=(i/2)[
\psi^+\alpha^a{\overrightarrow D_b} \psi-\psi^+{\overleftarrow
D}_b^+\alpha^a\psi ]$, and in the {\em artificial normal
coordinates}, discussed in Sec.\ref{sec:Sec3}, we had equation
(\ref{eq:E3.19}),
\begin{eqnarray}\label{eq:E6.1}
{\partial\over\partial x^\mu}\big[\sqrt{-g}~T^\mu_{~\nu} \big] = e
\sqrt{-g}j^\mu F_{\mu\nu}~.
\end{eqnarray}
As it should be, the kinematic acceleration
$w^\mu=e_{(0)}^\nu\nabla_\nu e_{(0)}^\mu$ of a charged particle
does not include a gravitational part (see \cite{Fock2}, \S 63).

The field $F_{\mu\nu}$ is a tensor and it satisfies the identity
(the first couple of Maxwell equations),
\begin{equation}\label{eq:E6.2}
F_{\mu\sigma\nu}\equiv\nabla_\mu F_{\sigma\nu}+
 \nabla_\sigma F_{\nu\mu}  + \nabla_\nu F_{\mu\sigma}=0.
\end{equation}
The divergence of this tensor, $\nabla^\mu F_{\mu\sigma\nu}=0$,
can be cast in the following form,
\begin{eqnarray}\label{eq:E6.3}
-\Box F_{\mu\nu} + R^\kappa_{~\mu} F_{\kappa\nu}- R^\kappa_{~\nu}
F_{\kappa\mu} - R_{\mu\nu\kappa\sigma}F^{\kappa\sigma}  \nonumber
\\ =( \nabla_\mu [\nabla^\sigma F_{\sigma\nu}] - \nabla_\nu
[\nabla^\sigma F_{\sigma\mu}])~,~~~~~~~
\end{eqnarray}
where $\nabla^\sigma=g^{\sigma\lambda}\nabla_\lambda$.  The l.h.s.
of this equation is the wave operator for the field $ F_{\mu\nu}$.
The two terms in the r.h.s. can be transformed further after we
postulate the second couple of Maxwell equations,
\begin{equation}\label{eq:E6.4}
\nabla_\mu F^{\mu\nu} =
eJ^\nu~,~~~~~~~~\nabla_\mu J^\mu=0,
\end{equation}
with the Dirac's vector current $eJ^\mu$ in the r.h.s. This
amounts to the second (in fact, independent) definition of
electric charge as the divergence of the electric field, and a few
reservations must be made. First, without a good reason, the same
coupling constant, as in the connection $\Gamma_a$, is postulated.
Only gauge invariance as an independent principle can provide for
an unquestionable equality of these constants. Second, the Dirac
field in Eq.(\ref{eq:E6.3}) is assumed to be a stable
configuration and the field $\aleph_\mu$ is considered a gradient.
Otherwise, the conservation of the vector current and the
non-conservation of the axial current  will conflict with Maxwell
equations. Third, the interaction between the electromagnetic
field and the spacelike axial current or pseudoscalar density,
which are present in Eq.(\ref{eq:B.10}), and affect the balance of
momenta inside the Dirac object, are disregarded. Only those
interactions that are responsible for the change of timelike
components of the momentum of a stable particle are allowed to be
sources of electromagnetic field. Then Eq.(\ref{eq:E6.2}) becomes,
\begin{eqnarray}\label{eq:E6.5}
-\Box F_{\mu\nu} + R^\kappa_{~\mu} F_{\kappa\nu}-
 R^\kappa_{~\nu}F_{\kappa\mu} -
R_{\mu\nu\kappa\sigma}F^{\kappa\sigma}=eQ_{\mu\nu}~, \nonumber \\
Q_{\mu\nu}=(\nabla_\mu J_\nu - \nabla_\nu J_\mu),~~~~~~~~~~~~
\end{eqnarray}
where $Q_{\mu\nu}$ is a convenient  intermediate notation.

The crucial question is, if $J_\mu$ in this equation is (or can
be) the current $j^\mu$ of Eq.(\ref{eq:E6.1}), which was derived
as a consequence of the Dirac equation with the potential $A_c$ in
the connection $\Gamma_c$. Evidently, the answer is {\em no}
because then, according to Eq.(\ref{eq:E4.1}), we must have
$Q_{\mu\nu}=0$. Therefore, an on-mass-shell object cannot be a
source of an electromagnetic field that may result in the Lorentz
force of self-interaction. The definition (\ref{eq:E6.4}) {\em
must} be complemented by Eq.(\ref{eq:E6.1}), which then determines
the measured acceleration of another charge that senses the field
of the first one. The potential $A_c$ in the Dirac equation that
describes a localized object must be  \textquotedblleft external";
its source can be only the current of another object. The problems
of mass and charge (including the problem of electromagnetic mass)
are not a single-body problem. One further implication of this
observation is that if one can simultaneously identify two
localized objects, then the Dirac fields of these objects cannot
overlap in space-time\footnote{From perspective of the second
quantization, when eigenfunctions of the Dirac equation are
associated with different states, this means that the Fock
operators of these two states must anti-commute!}. Since, for a
stable object, one cannot set $ej^\nu =\nabla_\mu F^{\mu\nu}$ in
the expression for the Lorentz force, it is also impossible to
express this force as the divergence of the energy-momentum
tensor, i.e., as  $~e j^\mu F_{\mu\nu}=\nabla_\sigma
F^{\sigma\mu}F_{\mu\nu}= -\nabla_\mu\Theta^\mu_\nu(F)$ and claim
that $\Theta^0_0$ is the energy density of the electromagnetic
field. This is not surprising since one cannot convert this energy
into any other form without a second object.

For an isolated stable charged object the condition that it cannot
interact with its own electric field means  that the only
 \textquotedblleft potential" in the wave equation (\ref{eq:E5.2}) is
$\aleph \propto 1/r^2$. The wave equation with such a steep
potential may have a strongly localized solution regardless of the
sign of this potential. The difference from the commonly studied
cases is that now the signs of this potential for the left- and
right-handed components are opposite and that, for a stable wave
form, the region $mr<1$ is cut off by Eq.(\ref{eq:E4.15}) as
unphysical. Therefore, a precursor of a localized state is present
in the Dirac equation even before the universal nonlinear
mechanism of the time slowdown takes over.

The field which is measured via  the Lorentz force (\ref{eq:E6.1})
always is a  \textquotedblleft field in vacuum". The wave equation
(\ref{eq:E6.5}) for the $F_{\mu\nu}$ from Eq.(\ref{eq:E6.1}) is a
homogeneous equation, which depends on the Dirac field of
(\ref{eq:E6.1}) only parametrically, via derivatives of
$g_{\mu\nu}(\psi)$ in the Riemann tensor.  The electromagnetic
sector of the theory turns out to be entirely in the form required
by Riemannian geometry. This sector is responsible for the
propagation of signals that are used to synchronize macroscopic
clocks (which is unambiguous only in special
relativity).\footnote{ It is important to emphasize that the
l.h.s. of Eq.(\ref{eq:E6.5}) is given in terms of measurable
electric and magnetic fields; therefore we indeed are dealing with
a signal that may have leading and rear fronts. Since the  world
time ${\cal T}$ (\ref{eq:E4.10}) is a harmonic function it can be
discontinuous along characteristics. The time of emission of a
photon, in principle, is not defined. A photon does not constitute
a signal.} A stable waveform of the Dirac field with
$Q_{\mu\nu}=0$ neither interacts with its own electromagnetic
field nor can it emit an electromagnetic field, as a signal, by
itself. This is yet further evidence that the object is in a
stationary state. This property is in line with the well known
fact that equation of the Coulomb's law is a constraint and not an
equation of motion. The longitudinal part of the electromagnetic
field does not propagate; the Coulomb field is simultaneous with
its source. The field of radiation emerges only when this
simultaneity is lost. This is yet another view of the realm of the
well-known phenomena of transient processes where the proper field
of a particle is truncated\cite{Feinberg}.

What if it occurred possible to trace an observed radiation field
back to the current in the interior of the localized object (so
that $J_\mu=j_\mu$) and, e.g. by a precise analysis of radiation,
to learn that $Q_{\mu\nu}\neq 0$ there? Then Eq.(\ref{eq:E4.1})
must be replaced by the second equation of (\ref{eq:E6.5}).
Proceeding as previously, we get
\begin{eqnarray}\label{eq:E6.6}
\nabla_\mu e^{(0)}_\nu - \nabla_\nu e^{(0)}_\mu  +
e^{(0)}_\nu\partial_\mu\ln{\cal R} -
e^{(0)}_\mu\partial_\nu\ln{\cal R}  \nonumber\\ =-(1/{\cal R}
)Q_{\mu\nu}. ~~~~~~~~~~~~~
\end{eqnarray}
Contracting this equation with spacelike $ e_{(i)}^\nu
e_{(j)}^\mu$ ($i,j=1,2,3$) and recalling Eqs.(\ref{eq:E2.19})  we
find that
\begin{equation}\label{eq:E6.7}
\omega_{0ij}-\omega_{0ji} =-(1/{\cal R} )Q_{ij}~.
\end{equation}
If $Q_{ij}\neq 0$ starting from some time moment $t_0$,  then at
$t>t_0$ the congruence $e_{(0)}^\mu$ of lines of the vector
current cannot be a normal congruence. The family of spacelike
surfaces $t=const$, orthogonal to the vector current, vanishes.
This means that Eq.(\ref{eq:E4.4}) cannot be obtained and Dirac
field cannot form a stable object\footnote{\label{fn:Meiss}This is
yet another way to view two seemingly different phenomena,  the
Meissner effect and precession of the spin in magnetic field. In
order to be in a stable quantum state, the superconductor expels
magnetic field from its interior or confines it into vortices,
thus defining a common time across its whole volume. In the same
way, electron with the magnetic moment (\ref{eq:E4.28}), being
placed in magnetic field, moves in precession with the Larmor
frequency. Therefore, in rotating frame the magnetic field
vanishes and the electron still can have the same world time
across its volume (staying in a certain quantum state). }. The
electromagnetic fields produced by such an object are not just
longitudinal (Coulomb) fields and the object must start to radiate
solely because the electromagnetic field around it is not
simultaneous with its source. Since the Dirac equation is of the
hyperbolic type, the changes of the Dirac field must propagate
also, having a light cone as a leading wave front.

Contracting Eq.(\ref{eq:E6.6}) with  $ e_{(0)}^\nu e_{(i)}^\mu$ we
obtain another equation,
\begin{equation}\label{eq:E6.8}
\omega_{i00}=-(\partial\ln{\cal R}/ \partial s_i) -(1/{\cal R}
)Q_{i0},
\end{equation}
that accounts for the effect of the electric field, which  adds a
boost in the direction of the congruence  $e_{(i)}^\mu$.
Interaction with the electric field alone (which can be the case
only when this field is longitudinal) does not destroy the
hypersurfaces of constant time of a localized object, which allows
it to stay intact. The most important effect of acceleration in an
electric field is altering the shape of a charged object which
leads to an increase of its {\em internal energy} and of the local
charge density.

Referring to the above qualitative analysis  we may go further and
discuss a qualitative picture of some transient processes. If the
Dirac waveform is not stable (as is in the case of $\mu^+$) then
the development of instability (and the lifetime) must still be
stretched, due to the nonlinear effect of the time slowdown. When
the limit of stability (bifurcation) at $r\sim\lambda_\mu
=1.86\cdot 10^{-13} cm$ is reached, then the previous dynamic
regime suddenly breaks up and the field begins to evolve towards a
new configuration of a smaller mass $m_e$ and a larger
$\lambda_e=3.86\cdot 10^{-11} cm$. The Dirac field of the $\mu^+$,
which originally occupied the interval $\lambda_\mu<r<\lambda_e$,
must be radiated. Since the Dirac equation is hyperbolic, the
sharp front of the Dirac field, as any {\em precursor}, must
propagate along a characteristic (at the speed of light), in the
outward direction. This can only be the right component of the
Dirac spinor (which inherits its lightlike current from the
caustic), the ${\bar \nu}_\mu$. The final state of a $e^+$ emerges
not earlier than a new caustic at $ r = \lambda_e$ is formed. This
requires yet another transient process of the violent collapse of
the Dirac matter onto a new caustic and radiation of a similar
precursor, but with the opposite phase velocity, $\nu_e$. The
leptonic number is preserved dynamically in both processes.
Remarkably, it is exactly the existence of a well-defined (by the
spacelike {\em axial} vector) outward direction that eliminates
the illusion of the reflection symmetry of a plane wave and thus
predetermines a unique polarization of the spinor precursors. As
it was noted by Wigner \cite{Wigner2}, it is only a theoretical
idea of mirror symmetry (expressed in terms of {\em polar}
vectors) that hints of the possible existence of the second
polarization for lightlike spinor waveforms.

\section{ Conclusion. \label{sec:Sec7}}

The nonlinear Dirac equation, with its capricious interplay of the
many polarization degrees of freedom, poses a tough mathematical
challenge for theory. Its explicit solutions may well yield
various "magic numbers" that are currently known only from
experiment. Even before regular mathematical methods are
developed, one may rely on various qualitative consequences of the
finite size of the Dirac waveforms to re-analyze  existing data.

1. The conjectured connection between the mechanism of
self-localization and the sign of the electric charge of the Dirac
wave form also assumes that positively charged particles, which
are not perfectly stable,  must have a somewhat longer lifetime
than their negatively charged anti-particles. The ratio $\chi
=(\tau_+ -\tau_-)/\tau_{av}$ was measured for the most long-lived
species as a test of CPT-invariance. According to the Particle
Data Group, the difference in lifetime is indeed always positive,
$\chi(K^\pm)=(0.11\pm 0.09)\cdot 10^{-2}$, $\chi(\pi^\pm)=(5.5\pm
7.1)\cdot 10^{-4}$, $\chi(\mu^\pm)=(2\pm 11)\cdot 10^{-5}$, being
the largest for the heaviest specie.

2. One of the predicted manifestations of charge asymmetry is the
existence of the particle's external size. The internal radius is
universally limited from the below by the Compton length
$\lambda=\hbar/mc$. Due to the time slowdown in domains of large
Dirac density, the positively charged species must have smaller
external size than their negatively charged partners. Possibly,
$e^+$ has reasonably well defined external boundary, which then
may explain its relatively long lifetime in the environment of
normal matter. There may well exist observed differences in the
dynamics of electrons and positrons (or $p$ and ${\bar p}$, e.g.,
in accelerators) that are currently attributed to technical
issues.

3. There are certain coincidences of numbers that may prompt
another look at well known phenomena. We know that,
$\lambda_e=386fm$, $\lambda_\mu=1.86fm$. The radius of the proton,
as estimated via its electromagnetic form-factors, is $r_p=1fm$
and it grows for nuclei as $A^{1/3}$. This may well tell us
something about the high rate of $\mu^-$ capture by light nuclei
versus the low rate of the inverse $\beta$-decay by even heavy
nuclei. The correlation between the capture rate and size of a
nucleus may be a useful test.

4. The failure to keep anti-hydrogen molecule in the cold atom
trap for an indefinitely long time may establish the limits of
stability of the anti-proton in  anti-matter surroundings. One may
think of the capture of $e^+$ by a poorly localized ${\bar p}$ as
a possible mechanisms of instability.

5. The Lorentz contraction of the accelerated waveforms is a
dynamic effect, which leads to the accumulation of energy in an
extremely small volume and its release in the course of a
collision. The 5-8 Gev electrons and positrons are compressed to a
size about $10^{-1}fm$ having a density higher than a proton. Upon
colliding, they stop and create a sharp peak of invariant density,
which is very far from a stable configuration and rapidly decays.
The $B^\pm$ lifetime is reasonably long, $1.7\cdot 10^{-12}s$, and
its size is about $10^{-1}fm$. The time slowdown at ${\cal R}\gg
1$ may result in two effects: (i) an abnormally long lifetime of
the resonance and (ii) an exotic trend to further decay into
compact heavy objects rather than to decay into lighter objects
according to the usual spectator model. The mode $B^+\to K^+
X(3872)\to J/\psi \pi^+\pi^-$ seems to be a candidate for this
kind of the process because the width of $X(3872)$ is very small.

\appendix

\section{ Parallel transport of Dirac field.
\label{app:appA}}
\renewcommand{\theequation}{A.\arabic{equation}}
\setcounter{equation}{0}

In order to derive a measure for comparison of the fields
$\psi(x_1)$ and $\psi(x_2)$ at two close points let us require,
following Fock \cite{Fock1}, that the components $j^{a} =\psi^+
\alpha^a \psi $ are the invariants of the vector $j^\mu(x)$ and
the congruences $e_{(a)}^\mu(x)$ at the point where vector is
defined, $j^{a}=e^{(a)}_\mu j^\mu $. For now, we assume that a set
of four orthogonal congruences is fixed in advance and that Dirac
matrices $\alpha^a$ are either invariants or covariantly constant
objects. When the invariant $j_a$ is parallel-transported by
$ds_b$ along an arc of congruence $(b)$, then, solely because the
local pyramid is being rotated, it must change by $~ \delta j_a=
\omega_{acb}j^c ds^b=\omega_{acb}\psi^+\alpha^c \psi ds^b~~$. The
invariants $\omega_{abc}$ are defined by Eq.(\ref{eq:E2.15}). Let
matrix $\Gamma_a$ (the  connection) define the change of the Dirac
field components in the course of the same infinitesimal
displacement, $\delta \psi= \Gamma_a \psi ds^a$, $~\delta
\psi^+=\psi^+ \Gamma^+_a ds^a~$. Let differential of the product
$\psi^+ \alpha^a \psi$ obey Leibnitz rule. This gives yet another
expression for $\delta j_a$,
\begin{equation}\label{eq:A.1}
\delta j_a= \psi^+ (\Gamma^+_b \alpha_a+\alpha_a\Gamma_b)\psi
ds^b~.
\end{equation}
The two forms of $\delta j_a$ must be identical. Hence, the
equation that defines $\Gamma_a$ is
\begin{equation}\label{eq:A.2}
\Gamma^+_b \alpha_a+\alpha_a\Gamma_b=\omega_{acb} \alpha^c ~,
\end{equation}
and it has the most general solution,
\begin{eqnarray}\label{eq:A.3}
\Gamma_b(x)=ieA_b(x)+ig\rho_3 \aleph_b(x) \nonumber \\* - {1\over
2}\omega_{0kb}(x)\rho_3\sigma_k -{i\over
4}\epsilon_{0kim}\omega_{imb}(x)\sigma_k~,
\end{eqnarray}
where the last two terms can be compacted as,
$\Omega_b=(1/4)\omega_{cdb}\rho_1\alpha^c\rho_1\alpha^d=
(1/4)\omega_{cdb}\gamma^c\gamma^d~$. These two terms correspond to
an infinitesimal boost of (\ref{eq:E2.4}) along the spatial
$k$-axis with parameter $\omega_{0kb}ds^b$ and an infinitesimal
rotation of (\ref{eq:E2.3}) in the $(im)$-plane  with parameter
$\omega_{imb}ds^b$, respectively. This analogy, however, is
limited. While Eqs.(\ref{eq:A.1})-(\ref{eq:A.3}) do imply some
measure for the length of an arc (and of an angle as the ratio of
the two lengths), Eq.(\ref{eq:E2.2}) does not. The first two terms
are due to an intrinsic indeterminacy that arises when one has to
compare Dirac fields at two different points relying only on the
properties of the vector forms $e_{(a)}^\mu(\psi)$. The first term
is readily associated with the electromagnetic potential. The
second one would not appear at all if, following Fock
\cite{Fock1}, we required that $\delta(\psi^+\rho_1\psi)=0$ and
$\delta(\psi^+\rho_2\psi)=0$. This decision was motivated by that
kind of invariance of the Dirac equation in Minkowski space (local
Lorentz invariance),  which is not inherited by the Dirac field in
Riemannian geometry. The position of  $\aleph_b$ in connection
(\ref{eq:A.3}) may lead to the impression that it can well be a
 \textquotedblleft next field", which interacts with the axial
current ${\cal J}_b$ of the Dirac field and is governed by an
independent equation of motion.  At least for the on-mass-shell
Dirac field, this is not true.

The connection (\ref{eq:A.3}) commutes with the matrix $\rho_3$ so
that Eq.(\ref{eq:A.2}) remains the same when
$\alpha_a\to\rho_3\alpha_a$. It neither commutes nor anti-commutes
with $\rho_1$ and  $\rho_2$, viz.
\begin{eqnarray}\label{eq:A.4}
\Gamma^+_b \rho_1+\rho_1\Gamma_b= 2g\rho_2 \aleph_b ~,~~\nonumber
\\ \Gamma^+_b\rho_2+\rho_2\Gamma_b= -2g\rho_1 \aleph_b ~.~
\end{eqnarray}

From now on, we postulate that invariant derivative of the Dirac
field is $D_a\psi=(\partial_a-\Gamma_a)\psi$ where $\partial_a=
e_a^\mu\partial_\mu$ is the derivative in the direction of a curve
of congruence $(a)$. Assuming the Leibnitz rule for $D_a$ and
considering all Dirac matrices as constants we readily reproduce
the reference point of Eqs.(\ref{eq:A.1}) and (\ref{eq:A.2}) as
\begin{eqnarray}\label{eq:A.5}
D_bj_a= \partial_b j_a -\omega_{acb} j_c \equiv \nabla_b j_a ~.
\end{eqnarray}
The result (\ref{eq:A.5}) for $D_bj_a$ is a warrant that after
projecting the r.h.s. into coordinate space we must recover the
covariant derivative $\nabla_\mu j_\nu$. Indeed, $ e^a_\mu e^b_\nu
\nabla_b j_a =\partial_\mu j_\nu - \Gamma^\sigma_{\nu\mu} j_\sigma
=\nabla_\mu j_\nu ~$, and we shall consider this  as a proof that
$j_a$ is an invariant of the vector $j_\mu$ and congruence
$e^\mu_{(a)}$\footnote{Schouten (\cite{Schouten}, Ch.II, \S 9;
Ch.III, \S 9) considers equations like (\ref{eq:A.5}) as a
condition that fixes the components of a vector with respect to
anholonomic coordinate system.}. (There is no one-to-one
correspondence between the two terms of $\nabla_bj_a$ and
$\nabla_\mu j_\nu$!)

In exactly the same way we may verify that the invariants $D_b
{\cal J}_a$ of the axial current are of the form $ D_b{\cal J}_a=
\nabla_b {\cal J}_a $ and conclude that $e^a_\mu e^b_\nu D_b{\cal
J}_a =\nabla_\mu {\cal J}_\nu$. {\em Vice versa}, the quantities
$D_b{\cal J}_a =e_a^\mu e_b^\nu \nabla_\mu{\cal J}_\nu$ are
invariants of a tensor and a system of congruences. It is
straightforward to verify (computing all derivatives as functions
of $\psi$) that equations like (\ref{eq:A.5}) hold not only for
$j^\mu$ and ${\cal J}^\mu$ but for $e_{(0)}^\mu[\psi]= j^\mu/{\cal
R}$ and  ${e_{(3)}^\mu[\psi]=\cal J}^\mu/{\cal R}$. Recalling
discussion of Sec.II, we may view this as complementary to
(\ref{eq:E2.14}), i.e., proof that the unit vectors $e_{(0)}^\mu$
and $e_{(3)}^\mu$ are  vectors of Riemannian geometry.  For the
same reason, the projectors $[\delta^\mu_\nu-j^\mu j_\nu/{\cal
R}^2]$ and $[\delta^\mu_\nu+ {\cal J}^\mu {\cal J}_\nu/{\cal R}^2]
$ are the tensors.

Using the same technique of differentiating and by virtue of
Eqs.(\ref{eq:A.4}) we obtain
\begin{eqnarray}\label{eq:A.6}
D_a{\cal S}=\partial_a{\cal S} -2g{\cal P} \aleph_a,~ D_a{\cal
P}=\partial_a{\cal P} + 2g{\cal S} \aleph_a.
\end{eqnarray}
As one can see, that there is no immediate correspondence between
the algebraic and differential properties of the scalars. However,
the quantities from  the first line of (\ref{eq:E2.8}) (like
${\cal R}$) are differentiated as true scalars. The same behavior
is observed for the components of the skew-symmetric tensor ${\cal
M}_{ab}$. Instead of the anticipated $e^c_\lambda e^a_\mu e^b_\nu
D_c {\cal M}_{ab}=\nabla_\lambda{\cal M}_{\mu\nu}$  we  encounter
one more disagreement with the differential criterion
(\ref{eq:A.5}),
\begin{eqnarray}\label{eq:A.7}
D_c {\cal M}_{ab}= \nabla_c{\cal M}_{ab} -2g \aleph_c
{\overstar{\cal M}}_{ab},\\ D_c {\overstar{\cal M}}_{ab}=
 \nabla_c {\overstar {\cal M}}_{ab} +2g \aleph_c {\cal M}_{ab}.\nonumber
\end{eqnarray}
We leave open the question of  if and when  $e_{(1)}^\mu$ and
$e_{(2)}^\mu$ of Eqs.(\ref{eq:E2.7}) are the vectors with the same
degree of confidence as  $e_{(0)}^\mu$ and $e_{(3)}^\mu$. In most
cases, $e_{(1)}^\mu$ and $e_{(2)}^\mu$ correspond to angular
coordinates, which are physically uncertain without external
fields (other objects nearby). The additional terms in
Eqs.(\ref{eq:A.7}) indicate that these congruences can be normal
only in very rare cases of solutions with a dynamically generated
symmetry. The expected one-to-one match with geometry is spoiled
by the extra (with respect to generator of Lorentz transformations
in the connection $\Gamma_a $) matrices $\rho_1$ and $\rho_2$
responsible for the  \textquotedblleft mixing" between right and
left components. From this perspective, the Sakharov's idea
\cite{Sakharov1}, regarding the topological nature of elementary
charges, seems be closer to reality that it initially appeared. A
further analysis of this issue seems to be impossible without an
explicit solution of the equations of motion.

Recalling Eqs.(\ref{eq:E2.9}) and using (\ref{eq:A.6})  we can
compare the covariant derivative $D_a{\cal P}$ computed in two
ways, as $D_a{\cal P}=\partial_a({\cal R}\cdot\sin\Upsilon) +
2g{\cal R} \cos\Upsilon\aleph_a$ or, alternatively, as $D_a{\cal
P}=\partial_a{\cal R}\cdot\sin\Upsilon + {\cal R} \cos\Upsilon
D_a\Upsilon$. The result reads as
\begin{eqnarray}\label{eq:A.8}
D_a\Upsilon[\psi]= \partial_a\Upsilon + 2g\aleph_a,
\end{eqnarray}
which is invariant under the simultaneous transformations,
$\Upsilon\to\Upsilon+Y(x)$ and $2g\aleph_a \to 2g\aleph_a
-\partial_a Y(x)$. A supposed freedom of such (chiral)
transformations is not permissible, since these transformations
change the observables in the r.h.s. of Eqs.(\ref{eq:E3.15}) and
(\ref{eq:B.4}) without altering the l.h.s.

The commutator $[D_a,D_b]$  still contains derivatives. Indeed,
\begin{eqnarray}\label{eq:A.9}
[D_a,D_b]\psi=(\partial_a\partial_b-
\partial_b\partial_a)\psi \nonumber\\
- [\partial_a\Gamma_b- \partial_b\Gamma_a-
\Gamma_a\Gamma_b+\Gamma_b\Gamma_a]\psi.
\end{eqnarray}
However, for practical purposes it is important that
$[D_a,D_b]\psi$ can be split into two parts, with and without
derivatives. Since $\psi$ is a coordinate scalar and, in general,
derivatives along arcs do not commute, we have $(\partial_a
\partial_b-\partial_b
\partial_a)\psi= (\omega_{cab}-\omega_{cba})\partial_c\psi$. Now
we can re-assemble $[D_a,D_b]\psi$ as follows,
\begin{eqnarray}\label{eq:A.10}
 [{\overrightarrow  D}_a,{\overrightarrow D}_b ]
\psi= (\omega_{cab}-\omega_{cba}) {\overrightarrow
D}_c\psi-\mathbb{D}_{ab}, ~~~~~~~\\ \mathbb{D}_{ab}=
[\partial_a\Gamma_b-\partial_b\Gamma_a-
\Gamma_a\Gamma_b+\Gamma_b\Gamma_a - C_{cab}\Gamma_c]\psi,\nonumber
\end{eqnarray}
where the term
$C_{cab}\Gamma_c\equiv(\omega_{cab}-\omega_{cba})\Gamma_c$  is
added and subtracted to replace $\partial_c\psi$ by the covariant
derivative $D_c\psi$.  The matrix operator
$\mathbb{D}_{ab}=-e_a^\mu e_b^\nu [D_\mu,D_\nu]$ in the r.h.s.
does not contain derivatives and can be explicitly found,
\begin{eqnarray}\label{eq:A.11}
\mathbb{D}_{ab}=-{1\over 4} R_{abcd}\rho_1\alpha^c\rho_1\alpha^d
+ieF_{ab}+ig\rho_3U_{ab},\\* F_{ab}=\partial_a A_b- \partial_b A_a
-(\omega_{cab}-\omega_{cba}) A_c,\nonumber\\*
 U_{ab}=\partial_a \aleph_b- \partial_b\aleph _a -
 (\omega_{cab}-\omega_{cba})\aleph_c,\nonumber
\end{eqnarray}
where invariants of the Riemann curvature tensor are defined by
Eq.(\ref{eq:E2.23}) and $F_{ab}$ and $U_{ab}$ are invariants of
the electromagnetic tensor $F_{\mu\nu}=\nabla_\mu A_\nu-\nabla_\nu
A_\mu$ and the field tensor  $U_{\mu\nu}=\nabla_\mu \aleph_\nu-
\nabla_\nu \aleph_\mu$, respectively.

Using the cyclic symmetry of the Riemannian curvature tensor it is
straightforward to show \cite{Fock1} that
\begin{eqnarray}\label{eq:A.12}
\alpha^a \mathbb{D}_{ab}={1\over 2} \alpha^a R_{ab} +ie\alpha^a
F_{ab}+ig\rho_3\alpha^a U_{ab}.
\end{eqnarray}
When field $\aleph_a$ in the connection $\Gamma_a$ is a gradient,
the tensor $U_{\mu\nu}$ vanishes identically, which is assumed
throughout this paper except for Eq.(\ref{eq:B.10}), where a
possible connection with electro-weak theory is mentioned.

\begin{widetext}

\section{Stress tensor and pion field.\label{app:appB}}
\renewcommand{\theequation}{B.\arabic{equation}}
\setcounter{equation}{0}

\subsection{Internal flux of mass and stress in the Dirac field.}

In this section we study the stress tensor $P^a_b=i\psi^+\rho_3
\alpha^a D_b \psi$, mostly following the same logic as for the
energy momentum tensor $T^a_b=i\psi^+\alpha^a D_b \psi$ in
Sec.\ref{sec:Sec3}B, starting from its covariant derivative. We
find that
\begin{eqnarray}\label{eq:B.1}
D_c[\psi^+ \rho_3\alpha^a {\overrightarrow D}_b \psi]=
 \partial_c[\psi^+ \alpha^a {\overrightarrow D}_b
\psi]- \omega_{adc}\psi^+\rho_3 \alpha^d {\overrightarrow D}_b
\psi.
\end{eqnarray}
Once again, the last term of Eq.(\ref{eq:E3.5}) is missing, and
thus we have no confidence that the covariant derivative is a
tensor. This time, let us begin by contracting indices $a$ and $b$
in Eq.(\ref{eq:B.1}),
\begin{eqnarray}\label{eq:B.2}
D_c[\psi^+\rho_3 \alpha^a {\overrightarrow D}_a \psi]=
\partial_c[\psi^+\rho_3 \alpha^a {\overrightarrow D}_a \psi]- \omega_{abc}\psi^+
\rho_3\alpha^b {\overrightarrow D}_a \psi.
\end{eqnarray}
By virtue of the Dirac equations, the first term in the r.h.s. of
(\ref{eq:B.2}) becomes $\partial_c[m\psi^+\rho_2\psi]$.
Alternatively, we can immediately use the equations of motion in
the l.h.s. and only then differentiate (matrices $\rho_3$ and
$\alpha^a$ commute),
\begin{eqnarray}\label{eq:B.3}
D_c[\psi^+ \rho_3\alpha^a {\overrightarrow D}_a \psi]= m
D_c[\psi^+\rho_2 \psi] = m\partial_c[\psi^+\rho_2 \psi]+ m \cdot
2g{\cal S}\aleph_c ~.
\end{eqnarray}
Comparing the last two equations we finally get the equation,
\begin{eqnarray}\label{eq:B.4}
\omega_{acb}\cdot P_{ca} =-2ig m {\cal S}\aleph_b ,
\end{eqnarray}
which is complementary to Eq.(\ref{eq:E3.15}). The imaginary part
in the l.h.s. is due to $(1/2)[P_{ca}-P^+_{ca}]=(i/2)D_c{\cal
J}_a$. Since the axial current is a vector, we can rewrite the
last equation as
\begin{eqnarray}\label{eq:B.5}
(1/2)\omega_{acb}\nabla_c{\cal J}_a =-2g m {\cal S}\aleph_b ,
\end{eqnarray}
which is complementary (dual) to Eq.(\ref{eq:E4.18}). The
skew-symmetric Hermitian part, $(P_{ca}+P^+_{ca})
-(P_{ac}+P^+_{ac})$, must vanish since the r.h.s. of
Eq.(\ref{eq:B.4}) is an imaginary quantity. This yields the
equation, which duplicates Eq.~(\ref{eq:E4.1}),
\begin{eqnarray}\label{eq:B.6}
i[\psi^+\rho_3 \alpha_a {\overrightarrow D}_c \psi - \psi^+
{\overleftarrow D}^+_c\alpha_a\rho_3\psi- \psi^+\rho_3 \alpha_c
{\overrightarrow D}_a \psi + \psi^+ {\overleftarrow
D}^+_a\alpha_c\rho_3  \psi] =\epsilon_{acut}D_uj_t=0.
\end{eqnarray}
and thus indicates that we still are dealing with a stable
waveform.

Contracting (in Eq.(\ref{eq:B.2})) indices $a$ and $c$ we arrive
at the expression, which is similar to Eq.(\ref{eq:E3.8}),
\begin{eqnarray}\label{eq:B.7}
D_a[\psi^+\rho_3 \alpha^a {\overrightarrow D}_b \psi]=
\partial_a[\psi^+\rho_3 \alpha^a {\overrightarrow D}_b \psi]+
\omega_{acc}\psi^+\rho_3 \alpha^a {\overrightarrow D}_b  \psi =
{1\over\sqrt{-g}} {\partial\over\partial x^\nu}\bigg[\sqrt{-g}
e^\nu_{(a)} ~(\psi^+\rho_3 \alpha^a {\overrightarrow D}_b \psi)
\bigg].
\end{eqnarray}
 Let us first rewrite the l.h.s. of Eq.~(\ref{eq:B.7}) as
\begin{eqnarray}\label{eq:B.8}
D_a[\psi^+ \rho_3\alpha^a {\overrightarrow D}_b \psi]=\psi^+
\rho_3\alpha^a [{\overrightarrow D}_a {\overrightarrow D}_b   -
{\overrightarrow D}_b {\overrightarrow D}_a] \psi +
\psi^+{\overleftarrow D}^+_a\rho_3\alpha^a{\overrightarrow D}_b
\psi  + D_b(\psi^+ \rho_3\alpha^a {\overrightarrow D}_a \psi) -
\psi^+{\overleftarrow D}^+_b\rho_3\alpha^a{\overrightarrow D}_a
\psi.~~
\end{eqnarray}
Because of an obvious change of signs (caused by an extra
$\rho_3$), the last three terms of (\ref{eq:B.8}) do not cancel.
Instead of (\ref{eq:E3.9}) we have
\begin{eqnarray}\label{eq:B.9}
D_aP^a_b =i\psi^+ \rho_3\alpha^a [{\overrightarrow D}_a
{\overrightarrow D}_b -{\overrightarrow D}_b {\overrightarrow
D}_a] \psi +im D_b{\cal P} +im[\psi^+ \rho_2{\overrightarrow
D}_b\psi-\psi^+{\overleftarrow D}^+_b \rho_2\psi].
\end{eqnarray}
By splitting the commutator according to (\ref{eq:A.10}), we can
assemble the covariant derivative in the l.h.s. as
\begin{eqnarray}
\nabla_\mu P^\mu_\nu=\nabla_\mu {\sf Re}(P^\mu_\nu) +{i\over 2}
\nabla_\mu\nabla_\nu{\cal J}^\mu=\nabla_\mu {\sf Re}(P^\mu_\nu)
 +{i\over 2}{\cal J}^\mu R_{\mu\nu}+im\partial_\nu{\cal P},\nonumber
\end{eqnarray}
leaving on the r.h.s a remainder, $ e^b_\nu[\omega_{cab}P^a_c +
e{\cal J}^a F_{ab}+gj^a U_{ab}+(i/2){\cal J}^a R_{ab}]$. By virtue
of Eqs.(\ref{eq:B.4}) and (\ref{eq:A.6}) the imaginary terms on
both sides exactly cancel each other  and the remaining real part
reads as
\begin{eqnarray}\label{eq:B.10}
\nabla_\mu {\sf Re}(P^\mu_\nu)=e{\cal J}^\mu F_{\mu\nu} +gj^\mu
U_{\mu\nu} +im[\psi^+ \rho_2{\overrightarrow
D}_\nu\psi-\psi^+{\overleftarrow D}^+_\nu \rho_2\psi].
\end{eqnarray}
\end{widetext}
The flux of momentum in the spacelike direction is determined by
the Lorentz force that acts on the {\em electric axial current}
$e{\cal J}^\mu$, but also by the Lorentz force of the field
$\aleph_\mu$ that acts on the vector current $gj^\mu$. This part
is logical to write down in terms of the mixed fields that act on
lightlike left and right Dirac currents. It describes the
interplay between the right and left components of a Dirac
particle and belongs to the area of electro-weak interactions.
Being interested only in stable states, we do not consider it in
any details here. The last term is due to convection transport of
the pseudoscalar mass density $m{\cal P}$ {\em in an
electromagnetic field}. An importance of axial electric forces for
the processes of pion electro-production was noticed by Nambu and
Shrauner \cite{Nambu1}. The convection term can be cast as $$
m[\psi^+ \rho_2(i\partial_b\psi) -(i\partial_b\psi^+)\rho_2\psi
+2eA_b{\cal P} +(1/2)\omega_{cdb} {\overstar{\cal M}}^{cd}],$$
which repeats the familiar pattern of Gordon's decomposition of
the Dirac (vector) current with the replacement $\rho_1\to\rho_2$,
$e\to m$, and where the long derivative includes {\em only the
electromagnetic potential}. The pseudoscalar density ${\cal P}$ is
one of many polarization degrees of freedom of the Dirac field and
is not an independent field. In some approximation, the charged
pseudoscalar flux inside Dirac waveforms (e.g., nuclei) can be
viewed as the interaction of highly localized nucleons via soft
pion exchange. Complementary positions of vector and axial
currents in Eq.(\ref{eq:B.10}) prompt a parallel between the
charge and chirality of the Dirac field. This parallel was a point
of departure for the model of elementary particles developed by
Nambu and Jona-Lasinio \cite{Nambu2}.

Eq.(\ref{eq:B.10}) also describes a process within a compact
object that take place during a period of its acceleration. In its
course, the object changes configuration, undergoes Lorentz
contraction and, in fact, becomes a different object, with a
larger density and slower flowing time in its interior. Therefore,
the annoying non-unitarity of the proper Lorentz transformations
is a physical effect.

\subsection{Pion field and $\pi^0 \leftrightarrow 2\gamma$ processes.
\label{sapp:appB2}}
\renewcommand{\theequation}{B.\arabic{equation}}

The flux of charge, mass and momentum carried by the pseudoscalar
density in the interaction between Dirac nucleons is commonly
attributed to the pion field. Pions can also be detected as
sufficiently long lived particles. Therefore, ${\cal P}$ should
satisfy the Klein-Gordon equation, which we will derive below.

The first step is to put the axial current in a form with
separated convection and polarization currents, as is done in
Gordon's decomposition of the vector current,
\begin{eqnarray}\label{eq:B.11}
{\cal J}^a=-{1\over 2m} \eta_{(a)} D_a {\cal P}+{1\over
2m}I^a;~~~~~ \\ I^a=-{1\over 2}[\psi^+\alpha^{[a}\rho_2\alpha^{b]}
{\overrightarrow D}_b\psi-\psi^+{\overleftarrow D}^+_b
\alpha^{[a}\rho_2\alpha^{b]} \psi].\nonumber
\end{eqnarray}
where $\alpha^{[a}...\alpha^{b]}$ stands for $\alpha^a...\alpha^b
- \alpha^b...\alpha^a$. (Now, the entire convection term is
reduced to the derivative of ${\cal P}$!) Computing the covariant
derivative of both sides of the last equation  and using
(\ref{eq:E3.4}) we obtain
\begin{eqnarray}\label{eq:B.12}
 D_a^2 {\cal P}=2\psi^+{\overleftarrow D}^+_a\rho_2
{\overrightarrow D}_a\psi - 2m^2{\cal P}\nonumber\\-{\sf Re}
[\psi^+\alpha^{[a}\rho_2\alpha^{b]}(D_aD_b-D_bD_a)\psi].
\end{eqnarray}
For the on-mass-shell Dirac field of the nucleon with a large mass
$m$, we may take the Dirac field in semi-classical approximation,
$\psi\propto e^{iS/\hbar}$, so that the first two terms in the
r.h.s. constitute the classical Hamilton-Jacobi equation for the
eikonal $S$. If this equation is satisfied with a sufficient
accuracy (the waveform behaves as a classical particle and the
resonance is sufficiently narrow), then in the r.h.s. remains only
the last term. By virtue of Eqs.(\ref{eq:A.10}) and
(\ref{eq:A.11}) this term becomes nothing but
$eF_{ab}\overstar{\cal M}^{ab}+({\mathsf R}_s/2){\cal P}
-(1/2)C_{cab}\psi^+\alpha^a\rho_2\alpha^b {\overrightarrow
D}_c\psi$, where ${\mathsf R}_s ={\mathsf R}[\psi^+,\psi]$ is the
scalar Riemannian curvature (with dimension $m^2$), which is a
functional of the Dirac waveform. As a result, we arrive at the
{\em independent} wave equation for the pseudoscalar density
${\cal P}$ (the pion field),
\begin{eqnarray}\label{eq:B.14}
 D_a^2 {\cal P}-{{\mathsf R}_s\over2}{\cal P}\approx
 -C_{cab}{\sf Re}[\psi^+\alpha^a\rho_2\alpha^b
{\overrightarrow D}_c\psi]\nonumber\\+
 eF_{ab}\overstar{\cal M}^{ab},~~~~~~~~~~~~~~~~~
\end{eqnarray}
a Heisenberg equation of motion with a {\em variable mass} defined
by the {\em negative} scalar Riemannian curvature ${\mathsf R}_s$
outside the stable nucleons \footnote{The negative curvature is
typical for the geometry of the expanding matter. The pions (and
mostly pions) are abundantly created in the high-energy collisions
where strong contraction of the colliding particles is translated
into the rapidity plateau in the distribution of pions. At any
given moment of time $t$, the proper time at  a distance $x$ from
a generic point $x=0$ corresponds to the earlier proper time
$\tau$ and has a larger density ${\cal R}(x,t)$. Therefore, as can
be perceived from any point, the proper time flows more slowly
with larger distance $x$ from this point.  As a result, Dirac
waves tend to refract into distant spatial regions. The issue of
stability of the expanding Dirac field and of its localization as
pions and muons will be discussed elsewhere.}. Quite surprisingly,
exactly {\em this} ${\cal P}$ enters the r.h.s. of
Eq.(\ref{eq:E3.20}) that defines the force of gravity/inertia in
the same approximation of a material point.

The source in the r.h.s. is Hermitian. Its first term is
\textquotedblleft geometric" and accounts for the flux of momentum
and twist of the tetrad basis which are needed in order that the
pion could be born. The second term is more related to the pion
decay and it can be rewritten in two ways. On the one hand it can
be presented as $2e\psi^+[\rho_1{\vec E}+\rho_2{\vec B}]\cdot{\vec
\sigma}\psi$. When $F_{ab}$ is the field of a standing transverse
electromagnetic wave it has a simple representation in terms of
the spin interaction with two waves of circular polarizations,
\begin{eqnarray}\label{eq:B.15}
 eF_{ab}\overstar{\cal M}^{ab}=4e \sum_k
 {-i\sqrt{\omega_k}\over\sqrt{2(2\pi)^3}}(C_k e^{-ikx}-C^*_k
e^{ikx}) \nonumber\\* \times [{\vec e}_L\cdot{\bar\psi}_R
{\vec\sigma}\psi_L +{\vec e}_R\cdot {\bar\psi}_L
{\vec\sigma}\psi_R ],~~~~~~~~~~
\end{eqnarray}
where ${\vec e}_{L,R}(k)\bot{\vec k}$ are the vectors of the two
circular polarizations, $\psi_{L,R}$ are the left and right
components of the Dirac spinor field, $\omega_k=E_\gamma$ is the
\textquotedblleft photon's energy" and $C_k$ is the Fourier
component of the initial or final (possibly, coherent) state of
the Heisenberg field $F_{ab}$. This form of the source of the pion
field allows one to qualify $\pi^0$ as a resonance in the system
of the Dirac field and a standing electromagnetic wave formed by
the two circular polarizations, which causes the simultaneous flip
of helicity of both components of the Dirac field. On the other
hand the source in the wave equation (\ref{eq:B.14}) has the
structure of the axial anomaly. This could be an exact
correspondence if there was a simple proportionality between
${\cal M}^{ab}$ and $F^{ab}$. Then, $eF_{ab}\overstar{\cal
M}^{ab}= CeF_{ab}\overstar{F}^{ab}$, where the explicit value of
$C$ must comply with the observed rate of the $\pi^0\to 2\gamma$
decay. It is instructive that the wave equation for the
pseudoscalar meson field ${\cal P}$ (that yields the pole in the
pion propagator) was derived exactly from the original equation
Eq.(\ref{eq:E3.4}). The term, which was {\em ad hoc} added to this
equation by S. Adler \cite{Adler} (in order to save the Ward
identity for the axial vertex in triangle graph) has naturally
appeared as the source in the wave equation. A uniform method to
derive wave equations for meson fields is beyond the scope of this
paper \cite{Me1}.

In the first approximation, the value of mass of the Dirac field
is not important. For this particular resonance, the mass term in
the l.h.s. of Eq.(\ref{eq:B.14}) can be confidently identified
with and measured as $(2E_\gamma)^2$. In fact, $m_\pi^2 \sim -
{\mathsf R}_s/2$ is an independent of the Dirac mass $m$ measure
of the \textquotedblleft metric elasticity" in the ground state of
the Dirac field, when balance between left and right is probed by
electromagnetic field, thus being a fundamental constant. The
geometry of currents inside $\pi^0$ as a finite-sized object is
not clear so far. The dynamic quantity $m_\pi$ is meaningful only
for $\approx \! 10^{-16} s$ of the resonance spike of the
pseudoscalar density. The form of the source (sink) in the
Eq.(\ref{eq:B.14}) indicates that this spike is a parametric
resonance which can be created in a reaction like $\gamma
p\to\pi^0 p \to \gamma\gamma p$ (or, more precisely,
$\gamma^*\gamma^*\to\pi^0\to \gamma\gamma $). It decays due to
axial polarization currents (eventually producing two photons of
the opposite circular polarization). The nucleon is needed solely
to trigger the process -- to locally determine the difference
between left and right or inward and outward. Another way to
excite the $\pi^0$ resonance is $e^+e^-$-annihilation at high
energy, etc. For the sake of completeness mention that the source
in Eq.(\ref{eq:B.14}) is complementary to the interaction of the
axial current with the electromagnetic field given by the term
$e{\cal J}^\mu F_{\mu\nu}$ in Eq.(\ref{eq:B.10}), which describes
the balance of momentum when the waveform changes its shape.

The processes with the charged pions can be considered in a
similar manner \cite{Me1}. The dynamic footing for the
phenomenological formalism of isospin symmetry, as it is presented
in various texts (e.g. \cite{Bjorken}), can be naturally built
from the basic properties of the Dirac field. Pions are identified
as one specific (pseudoscalar) polarization degree of freedom of
the Dirac waveforms. The totally dynamic  origin of the pion mass
term, which is determined by the curvature ${\mathsf R}_s
[\psi^+,\psi]$ of the matter-induced metric, explains the
diversity of faces that pions may reveal in different situations.
This variety ranges from soft pion glue within nuclei (when DIS
cannot resolve pion's structure functions) and up to free
propagation of the massive pions (tracks) at distances that allow
for the pion interferometry (which is sensitive to the microscopic
dynamics of the pions emission \cite{MS2}).

\begin{widetext}

\section{Angular variables in Dirac equation.\label{app:appC}}
\renewcommand{\theequation}{C.\arabic{equation}}
\setcounter{equation}{0}

Let us examine the properties of the solutions of the Dirac
equation (\ref{eq:E5.4}), neglecting nonlinear terms, in the
presence of the radial field $2g\aleph_r= -\partial_r \Upsilon$
and in a perfectly spherically symmetric geometry.  Since we will
focus on the nature of angular variables, an explicit dependence
$\aleph_r(r)$ is not essential. [One may think of
Eq.(\ref{eq:E4.17}) as an example.] The only non-vanishing
components of the Ricci rotation coefficients are $\omega_{212}=
(1/r)\cot\theta$ and $\omega_{131}=\omega_{232}=1/r$. Solely as a
reference, assume that there is an external electromagnetic
Coulomb field $A_0(r)$. Then the Dirac equation is
\begin{eqnarray}\label{eq:C.1}
\big[i\partial_0 -eA_0 +g\sigma_3\aleph_r
-i\rho_3\sigma_3(\partial_r +{1\over r}) 
-i{\rho_3\sigma_1 \over r}(\partial_\theta+{1\over 2}\cot\theta)
-i{\rho_3\sigma_2\over r\sin\theta}\partial_\varphi -m\rho_1
\big]\psi=0~.~~~
\end{eqnarray}
In terms of a new unknown function,
$~\tilde{\psi}(r,\theta,\varphi)=r\sqrt{\sin\theta}\psi~$, this
equation becomes
\begin{eqnarray}\label{eq:C.2}
 \big[i\partial_0 -eA_0 +g\sigma_3\aleph_r
+\rho_3\sigma_3(-i\partial_r)                  
+{\rho_3 \over r}(-i\sigma_1\partial_\theta-i
{\sigma_2\over\sin\theta}\partial_\varphi) -m\rho_1
\big]\tilde{\psi}=0~.
\end{eqnarray}

\end{widetext}

\renewcommand{\theequation}{C.\arabic{equation}}
Equation (\ref{eq:C.1}) is the Dirac equation in the tetrad basis.
In order to find its solution one has to separate the angular and
radial variables. This is known to be a somewhat tricky problem,
even in the standard problem with a radial Coulomb field(when
$\aleph_a=0~$). The Hermitian operators in Eq.~(\ref{eq:C.2}) are
the tetrad components of the momenta $p_3=-i\partial_r$,
$p_1=-ir^{-1}\partial_\theta$
$p_2=-i(r\sin\theta)^{-1}\partial_\varphi$. The operators $p_1$
and $p_2$ are clearly associated with the angular motion. If the
coefficients in this equation where not matrices, it would have
already been an equation with separated variables, which would
match the perfect spherical symmetry of the {\em external} fields
$A_0$ and $\aleph_r$. The problem is that the operators of the
radial and angular momenta do not commute (they anti-commute,
$[\alpha_3p_3,(\alpha_1 p_1+\alpha_2 p_2)]_+=0~$). A regular way
to avoid this obstacle is as follows \cite{Weyl1,Dirac1}: One
attempts to construct a minimal set of operators that commute with
the Hamiltonian. For example, one can check that the commutator
$[\alpha_3p_3,\rho_1(\alpha_1 p_1+\alpha_2 p_2)]_-=0~$ and take
the operator $\rho_1(\alpha_1 p_1+\alpha_2 p_2)$ as a generator of
the conserved quantum number. This trick works when $\aleph=0$ and
it is very instructive to see the details of its failure when
$\aleph\neq 0$.

The conventional operator of angular momentum is $\vec{\cal
L}=[\vec{r}\times\vec{p}]+\vec{\sigma}/2$. An additional operator
${\cal L}=\vec{\sigma}\cdot\vec{\cal L}-1/2$ commutes with the
orbital momentum, $[{\cal L},(\vec{r}\times\vec{p})]=0$, and has
the properties, ${\cal L}({\cal L}-1)=[\vec{r}\times\vec{p}]^2$
and ${\cal L}^2=\vec{\cal L}^2+1/4$. Therefore, if $\kappa$ is an
eigenvalue of operator ${\cal L}$ we obviously have
$\kappa(\kappa-1)=l(l+1)$ and $\kappa^2>0$. On the other hand, if
${\cal L}_A=\rho_1{\cal L}$, then $({\cal L}_A)^2={\cal L}^2$ and
these operators have the same sets of eigenvalues. In the tetrad
basis, these generators of the angular quantum numbers are
\begin{eqnarray}\label{eq:C.3}
{\cal L}_A=\rho_1 \big(-i\sigma_2\partial_\theta+ {i \sigma_1
\over\sin\theta}\partial_\varphi \big),~ {\cal L}_3=
-i\partial_\varphi +{1\over 2}\sigma_3.~
\end{eqnarray}
In terms of the {\em auxiliary} operator ${\cal L}_A$ (which {\em
has the same set of quantum numbers as the operator of the angular
momentum} but is a different operator) and the projection ${\cal
L}_3$ of angular momentum, the Dirac equation becomes
\begin{eqnarray}\label{eq:C.4}
[i\partial_0 -eA_0 +g\sigma_3\aleph_r
+\rho_3\sigma_3(-i\partial_r)  \nonumber\\*
 - \rho_2\sigma_3 {{\cal
L}_A \over r}-m\rho_1 ]\tilde{\psi}=0.
\end{eqnarray}
When the operator ${\cal L}_A$ commutes with all terms of
Hamiltonian (which is the case when $\aleph_\mu=0$) we can require
the wave function be an eigenfunction of the Hamiltonian and these
two operators,
\begin{eqnarray}\label{eq:C.5}
{\cal L}_A\tilde{\psi}=\kappa\tilde{\psi},~~~{\rm and}~~~ {\cal
L}_3\tilde{\psi}=(m_z+{1\over 2})\tilde{\psi}.
\end{eqnarray}
Since the ${\cal L}_A$ anti-commutes with $g\sigma_3\aleph_r$ we
have no obvious solution for the separation of variables.

Because the presence of the component $\aleph_r(r)$ at least
apparently preserves the spherical symmetry, we can try to look
for a general solution of the following form ($\xi$ and $\eta$ are
the left and right components of the Dirac field in spinor
representation, respectively),
\begin{eqnarray}\label{eq:C.6}
\tilde{\xi}=\!\!\left(\!\! \begin{array}{c} u_L(r,t) {\cal Y}
(\theta,\varphi) \\
 d_L(r,t){\cal Z}(\theta,\varphi) \end{array} \!\!\right)\!,
\tilde{\eta}=\!\!\left(\!\! \begin{array}{c}  u_R(r,t) {\cal Y}
(\theta,\varphi)\\
 d_R(r,t){\cal Z}(\theta,\varphi)  \end{array}\!\! \right).~
\end{eqnarray}
As a first step, we may try to substitute the Dirac spinor
(\ref{eq:C.6}) into Eqs.(\ref{eq:C.5}). One can immediately see
that the angular variables in (\ref{eq:C.5}) can be separated only
when $u_L= d_R$ and $u_R= d_L$. At the same time, by inspection of
the complete system of four Dirac equations,
\begin{eqnarray}\label{eq:C.7}
(i\partial_0 -eA_0 -g\aleph_r-i\partial_r) u_L {\cal Y}
 \!\!= m u_R {\cal Y}+ d_L{i\Lambda_- \over r} {\cal Z},~~
\nonumber\\ (i\partial_0 -eA_0 +g\aleph_r+i\partial_r) d_L {\cal
Z}\!\! = m d_R {\cal Z} +u_L{i\Lambda_+\over r}{\cal Y},~~
\nonumber\\ (i\partial_0 -eA_0 -g\aleph_r+i\partial_r) u_R {\cal
Y}\!\! =  m u_L {\cal Y} -d_R{i\Lambda_- \over r} {\cal
Z},~~\nonumber\\ ( i\partial_0 -eA_0 +g\aleph_r-i\partial_r) d_R
{\cal Z} \!\! =m d_L {\cal Z}- u_R{i\Lambda_+ \over r} {\cal Y},~~
\end{eqnarray}
where
\begin{eqnarray}\label{eq:C.8}
\Lambda_\pm=(\partial_\theta\pm {i \over\sin\theta}
\partial_\varphi),
\end{eqnarray}
one can see that the condition $u_L= d_R$ and $u_R= d_L$ is
inconsistent with the presence of the field $\aleph_a$. The Dirac
equation breaks up into two systems of equations for only two
radial functions which are incompatible unless $\aleph_r\!=\!0$.

Nevertheless, just by inspection, one can see that the angular
functions ${\cal Y}_{k,m}(\theta,\varphi)$ and ${\cal
Z}_{k,m}(\theta,\varphi)$ that satisfy the  equations,
\begin{eqnarray}\label{eq:C.9}
\Lambda_-{\cal Z}_{k,m}(\theta,\varphi)= -k {\cal
Y}_{k,m}(\theta,\varphi), \nonumber\\* \Lambda_+ {\cal
Y}_{k,m}(\theta,\varphi)= k {\cal Z}_{k,m}(\theta,\varphi),~~
\end{eqnarray}
do separate the angular variables in the Dirac equation (and do
not separate them in Eq.~(\ref{eq:C.5})). With this separation of
angular variables, Eqs.~(\ref{eq:C.7}) yield a system of four
differential equations for the four radial functions.
Eqs.~(\ref{eq:C.9}) are the equations for spherical harmonics but
$\theta$ and $\varphi$ are not the angles of the spatial angular
coordinates.


\begin{thebibliography}{99}

\bibitem{Wigner1} E. Wigner, Helv. Phys. Acta, Suppl., IV, 210 (1956 )
\bibitem{Sakharov1}  A.D. Sakharov, The Topological Structure of Elementary
Charges and CPT Symmetry, in Collected Scientific Works, p.199,
Marcel Dekker Inc.
\bibitem{Fock1}   V. Fock, Z. f. Phys. {\bf 57}, 261 (1929);
V. Fock and D. Ivanenko, Z. f. Phys. {\bf 54},  798 (1929).
\bibitem{Weyl} H. Weyl,  Z. f. Phys. {\bf 56},  330 (1929).
\bibitem{Cartan}  E. Cartan, The theory of spinors, Hermann, Paris.
\bibitem{Neeman}  Y. Ne'eman, Annales de l'I.H.P., sec. A, {\bf 28}, 369 (1978);
F.W Hehl, E.A Lord, Y. Ne'eman, Phys. Rev. D, {\bf 17}, 428
(1978).
\bibitem{Schouten} J.A. Schouten, Ricci-Calculus, Springer-Verlag, 1954.
\bibitem{Eisenhart} L.P. Eisenhart, Riemannian geometry, Princeton, 1926.
\bibitem{TLC} T.Levi-Civita, The absolute differential calculus,
London, 1926 (Dover, 1977) .
\bibitem{Wigner2} E. Wigner, Rev. Mod. Phys., {\bf 29}, 255
(1957).
\bibitem{Feinberg} E.L. Feinberg,  Sov. Phys. Usp., {\bf 23}, 629
 (1980).
\bibitem{Fock2}   V.A. Fock, The theory of space, time and gravitation, Macmillan, NY, 1964.
\bibitem{MS1} A. Makhlin, E. Surdutovich, Phys. Rev. C, {\bf 58}, 389 (1998).
\bibitem{Arnold} V.I. Arnold, Singularities of Caustics and Wave
Fronts, Springer, 1990; V.I. Arnold, B.A. Khesin, Topological
Methods in Hydrodynamics, Springer, 1999.
\bibitem{Nambu1} Y.Nambu, E. Shrauner, Phys. Rev., {\bf 128}, 862
(1962).
\bibitem{Nambu2} Y.Nambu, G. Jona-Lasinio, Phys. Rev., {\bf 122},
345 (1961).
\bibitem{Weyl1} H.Weyl, The theory of groups and quantum mechanics,
Dover Publications, 1931.
\bibitem{Dirac1} P.A.M. Dirac, The principles of quantum mechanics,
Oxford, Clarendon Press, 1958.
\bibitem{Adler} S. Adler, Phys. Rev., {\bf 177}, 2426 (1969).
\bibitem{Me1} In preparation.

\bibitem{Bjorken} J.D. Bjorken, S.D. Drell, Relativistic Quantum
Mechanics, McGraw-Hill, NY, 1965.
\bibitem{MS2} A. Makhlin, E. Surdutovich, Phys. Rev. C, {\bf 59}, 2761 (1999).



 \end{thebibliography}
 \end{document}